\documentclass[fleqn,usenatbib]{mnras}

\usepackage{amsmath}
\usepackage{amssymb}
\usepackage{subfig}
\usepackage{graphicx}
\usepackage{natbib}
\usepackage{soul}
\usepackage{xcolor}
\usepackage{caption}
\usepackage{adjustbox}

\usepackage[T1]{fontenc}
\usepackage{ae,aecompl}
\usepackage{newtxtext,newtxmath}

\newlength{\myfigwidtha}
\newlength{\myfigwidthb}
\title[Modelling CRE physics in hydro-simulation]{Modelling cosmic ray electron physics
in cosmological smoothed particle hydrodynamics simulation}

\author[Zheng~et~al.]{%
Dongchao Zheng $^{1}$
\thanks{Contact e-mail: dczheng21@outlook.com},
Weitian Li $^{1}$,
Zhenghao Zhu $^{1}$,
Chenxi Shan$^{1}$,
Jiajun Zhang$^{2}$
\newauthor
Linfeng Xiao $^{1}$,
Xiaoli Lian $^{1}$,
Dan Hu$^{1}$,
\\
$^{1}${School of Physics and Astronomy,
    Shanghai Jiao Tong University,
    800 Dongchuan Road, Shanghai 200240, China}\\
$^{2}${Center for Theoretical Physics of the Universe,
        Institute for Basic Science (IBS),
        Daejeon, 34126, Korea
    }\\
}

\date{Accepted XXX. Received YYY; in original form ZZZ}
\pubyear{2019}

\begin{document}
\label{firstpage}
\pagerange{\pageref{firstpage}--\pageref{lastpage}}
\maketitle

\begin{abstract}
Cosmic ray electron (CRE) acceleration and cooling are important
physical processes in astrophysics.
We develop an approximative framework to treat CRE physics
in the parallel smoothed particle hydrodynamics code Gadget-3.
In our methodology, the CRE spectrum of each fluid element is
approximated by a single power-law distribution
with spatially varying amplitude, upper cut-off, lower cut-off,
and spectral index. We consider diffusive shock acceleration to
be the source of injection, and oppositely the sinking processes
is attributed to synchrotron radiation, inverse Compton scatters,
and Coulomb scatters. The adiabatic gains and losses are also
included. We show that our formalism produces
the energy and pressure with an accuracy of $ > 90\%$
for a free cooling CRE spectrum. Both slope and intensity of
the radio emission computed from the CRE population given by
our method in cosmological hydro-simulation coincide well
with observations, and our results also show
that relaxed clusters have lower fluxes. Finally, we investigate 
several impacts of the CRE processes on the cosmological hydro-simulation,
we find that:
(1) the pressure of the CRE spectrum is very small and 
can be ignored in hydro-simulation,
(2) the impacts of the CRE processes on the gas phase-space state
of hydro-simulation is up to $3\%$,
(3) the CRE processes induce a $5\%$ influence on the mass function
in the mass range $10^{12} -10^{13} h^{-1} M_{\odot}$,
(4) The gas temperature of massive galaxy cluster 
    is influenced by the CRE processes up to $\sim 10\%$.
\end{abstract}



\section{Introduction}

Cosmological numerical simulation has become an indispensable tool in
studying of the structure formation process of the universe \citep{Millennium, Illustris}.
There are currently three techniques employed
in numerical simulations:
(1) grid-based Eulerian schemes with optional
adaptive mesh refinement \citep[AMR;][]{Enzo};
(2) particle-based Lagrangian methods, namely the smoothed particle hydrodynamics
\citep[SPH;][]{Monaghan_1992, Monaghan_2005, Gadget2, springel_2011};
(3) moving-mesh method
\citep{moving-mesh_2010, moving-mesh_2011}
improving on the weakness of the SPH and AMR.~In recent years,
with the rapid growth of computer performance and the implementation of more
sophisticated numerical algorithms, more complicated physical processes can be incorporated into
numerical simulations. For example, the most popular
TreeSPH code Gadget
\citep[%
\textbf{GA}laxies with \textbf{D}ark matter and
\textbf{G}as int\textbf{E}rac\textbf{T};
][]{Gadget2},
compared with its first version
\citep{Gadget1}, includes many
baryon physical processes, such as
star formation \citep{Springel_Hernquist_2003},
cosmic ray proton \citep[CRP;][]{Ensslin_2007,Jubelgas_2008},
cooling processes \citep{Scholz_1991, Katz_1996},
shock wave \citep{Pfrommer_2006},
thermal conduction \citep{Jubelgas_2004},
radiative transfer \citep{Petkova_2009},
magnetohydrodynamics \citep[MHD,][]{Dolag_Stasyszyn_2009},
black hole \citep{Blackhole1, Blackhole2}, and so on.

One of the major radiation mechanisms in the radio band is the synchrotron radiation of cosmic ray
electrons \citep[CRE,][]{Hoeft_Bruggen_2007}.
In order to trace the synchrotron radiation of CRE, we must know the spatial and energy
distribution of CRE as well as the magnetic field. Since the MHD simulation can infer the magnetic field, how to obtain the properties of CRE in simulation
is the key step to study the radio emission. The evolution of CRE spectrum is
described by the Fokker-Planck equation
\citep[FP;][]{Park_1995, Brunetti_2004, Pinzke_2017, Brunetti_2011},
which can be solved numerically  with the finite difference method
\citep{Chang_1970, Park_1996, Donnert_Brunetti_2014}.
However, solving the FP equation is both computation-intensive and memory-intensive.
It is inappropriate to apply the finite difference method to large-scale cosmic simulations directly.
Although \citet{Hoeft_Bruggen_2007} have proposed a novel method
to compress the data of CRE spectrum to reduce memory usage,
the computation-intensive problem still exists.
Post-processing of simulation data is another choice, which solves the FP equation over many simulation snapshots \citep{Pinzke_2017}.
But this scheme takes into account neither the simulation
information between snapshots nor the feedback of CRE physics.
In this work, by analysing the CRE
spectrum evolution governed by the FP equation,
we find that a power-law distribution,
characterised by
spatially varying amplitude, 
upper and lower cut-offs, 
and spectral index, 
is a good approximation for the practical CRE dynamics.
In such an approximate CRE framework, the intensive calculation
of numerically solving the FP equation can be avoided, which is
beneficial to investigating the CRE physical processes in
cosmological hydro-simulation.

This paper is organized as follows:
In Section~{\ref{sec:model}},
we describe our method to treat CRE physics in
cosmological hydro-simulation.
In Section~{\ref{sec:simu}},
we discuss the radio emission calculated from the CRE population and
the several impacts of the CRE processes
on cosmological hydro-simulation.
We conclude with a summary in Section~{\ref{sec:sum}}.

\section{Cosmic ray electron physics and modelling}\label{sec:model}

\subsection{Spectrum modelling of cosmic ray electron}\label{sec:pdf}
The population of relativistic particles injected by various astrophysical processes (e.g., AGN activities, SNe.) can be
approximated with a power-law distribution \citep{Hoeft_Bruggen_2007, Ensslin_2007, Jubelgas_2008}.
Therefore, we assume that the CRE spectrum in each
fluid element can be described by a single power-law with lower and upper cut-offs, i.e.
\begin{equation}
    \begin{split}
        f(p) &= 4\pi p^2 f(\textbf{p}) \\
        &= \frac{dN}{dpdV} = Cp^{-{\alpha}} \, \Theta(p-p_{\rm min}) \, \Theta(p_{\rm max}-p),
    \end{split}
    \label{eq:df}
\end{equation}
where the dimensionless momentum $p = |\textbf{p}|/{m_e c}$,
$\textbf{p}$ is electron momentum,
$m_e$ is the electron mass,
$c$ is the light speed,
$N$ is the number of electrons,
$V$ is the volume occupied by $N$ electrons,
$C$ is the normalisation,
$\alpha$ is the power-law slope,
$p_{\rm min}$ and $p_{\rm max}$ are the upper and lower cut-offs,
respectively, $\Theta$(x) denotes
the Heaviside step function.
If $p_{\rm max}$ is infinity and $p_{\rm min} \equiv q$, the number density $n$,
kinetic energy density $\epsilon$, pressure $P$, and average kinetic
energy $\overline{T} = {\epsilon}/{n}$ of CRE spectrum
are
(see appendix~\ref{sec:appd} for the detailed derivation):
\begin{subequations} \label{eq:n_e_P_T}
    \begin{equation}
        n(C, \alpha, q) = \int_{0}^{{\infty}}{dpf(p)} = \frac{C{q}^{1-{\alpha}}}{\alpha-1},
    \end{equation}
    \begin{equation}
        \begin{split}
            \epsilon(C, \alpha, q) =&\int_{0}^{{\infty}}f(p)T(p) dp = \frac{Cm_ec^2}{\alpha-1} \\
                \times \bigg[ \frac{1}{2} & B_{{1}/{(1+{q}^2)}}\left(\frac{\alpha-2}{2},
                \frac{3-{\alpha}}{2}\right) + {q}^{1-{\alpha}} \left(\sqrt{1+{q}^2}-1\right)\bigg],
        \end{split}
    \end{equation}
    \begin{equation}
        \begin{split}
               P(C, \alpha, q) =& \frac{m_e c^2}{3} \int_{0}^{{\infty}} f(p) \beta p dp \\
                           =& \frac{C m_e c^2}{6} B_{{1}/{(1+q^2)}}\left(\frac{\alpha-2}{2},
                           \frac{3-\alpha}{2}\right),
        \end{split}
    \end{equation}
    \begin{equation}
        \begin{split}
            \overline{T}(C, \alpha, q) =& \bigg[ \frac{{q}^{\alpha-1}}{2}B_{{1}/{(1+{q}^2)}}
                                   \left(\frac{\alpha-2}{2},\frac{3-\alpha}{2}\right) \\
                               +& \sqrt{1+{q}^2}-1\bigg]m_ec^2,
        \end{split}
    \end{equation}
\end{subequations}
where $T(p) = (\sqrt{1+p^2}-1)m_e c^2$ is the kinetic energy of a single electron
with momentum $p$, $\beta=v/c=p/\sqrt{1+p^2}$ is the dimensionless velocity, 
and $B_x(a,b) =\int_{0}^{x} t^{a-1} (1-t)^{b-1} dt$
denotes the incomplete Beta function.
Since there is no upper cut-off and
$p_{\rm min} > 0$, these equations are valid for $\alpha > 2$. For finite
$p_{\rm max}$, the values of $n, \epsilon$, $P$ and $\overline{T}$ can be derived from Eqs. \ref{eq:n_e_P_T}.

\subsection{Evolution of cosmic ray electron}

The temporal evolution of CRE distribution $f(p,t)$ is governed by the isotropic, gyro-phase averaged
FP equation (in the Lagrangian frame),
\begin{equation}
    \begin{split}
        \frac{df(p,t)}{dt} &= \frac{\partial}{\partial{p}} \left\{ f(p,t) \left[ \left| \frac{dp}
        {dt}\right|_{\rm cool}
        - \frac{1}{p^2} \frac {\partial}
        {\partial{p}}( p^2 D_{\rm pp} ) \right] \right\} \\
        &- (\nabla \cdotp v) f(p,t) + \frac{{\partial}^2}{\partial{p^2}} \left[ D_{\rm pp}f(p,t)\right]
        + Q\left[p,t; f(p,t)\right], \\
    \end{split}
    \label{eq:fp}
\end{equation}
where $Q$ is the injection function,
$d/dt = \partial/\partial{t} + v \cdotp \nabla $ is the Lagrangian derivative,
$v$ is the gas velocity,
the $\nabla \cdotp v$ represents adiabatic gains and losses,
$|{dp}/{dt}|_{\rm cool}$ represents Coulomb and radiative
losses including synchrotron radiation and inverse Compton scattering
\citep{Hoeft_Bruggen_2007,Longair_2011,Pinzke_2017},
which are given by:
\begin{subequations}
\begin{equation}
        \frac{dp}{dt}\bigg|_{\rm rad} = C_{\rm cool} p \sqrt{(1+p^2)},
    \label{eq:dpdt_rad}
\end{equation}
\begin{equation}
    \begin{split}
        \frac{dp}{dt}\bigg|_{\rm coul}  = &  \frac{3 c\sigma_{T} n_{\rm th, e}}{2\beta^2}
                \bigg\{\ln\left(\frac{m_ec^2\beta\sqrt{\gamma-1}}
                        {\hbar \omega_{\rm plasma}}\right) \\
                        &-\ln(2)\left(\frac{\beta^2}{2}+\frac{1}
                        {\gamma}\right) + \frac{1}{2} + \left(\frac{\gamma-1}{4\gamma}\right)^2\bigg\}, \\
    \end{split}
    \label{eq:dpdt_coul}
\end{equation}
with
\begin{equation}
        C_{\rm cool} = \frac{4 \sigma_T} {3 m_e c} \left[ (1+z)^4
        \frac{{B}_{\rm cmb,0}^2} {8 \pi} + \frac{B^2}{8\pi}\right]
\end{equation}
\begin{equation}
        \omega_{\rm plasma} = \sqrt{\frac{4\pi e^2 n_{\rm th,e}}{m_e}},
\end{equation}
where $\sigma_T$ is the Thomson cross-section, $B_{\rm cmb,0} \approx 3.24\,\mu \rm G$
is the equivalent magnetic field of the cosmic-microwave background at $z=0$, and $B$ is
the magnetic field. $\gamma=\sqrt{1+p^2}$ is the Lorentz factor, $\omega_{\rm plasma}$
is the plasma frequency, $\hbar$ is the reduced Planck constant, $n_{\rm th, e}$ is the number density of thermal electron, and
$e$ is the electron charge.
\end{subequations}
$D_{\rm pp}$ is the momentum space diffusion coefficient
\citep{ Brunetti_2004, Cassano_2005, Pinzke_2017},
which describes the turbulent acceleration. Compared to
the diffusive shock acceleration (DSA) investigated in this
work (see \ref{sec:inj}), the turbulent acceleration is 
relatively weak and inefficient, thus it is omitted in
this work (i.e. $D_{\rm pp} = 0$).

\subsection{Approximation method}\label{sec:met}
In this subsection, we explain the approximation methods
that are employed to determine the CRE spectrum
parameters (i.e. $C, \alpha, p_{\rm min}, p_{\rm max}$ in Eqs.~\ref{eq:n_e_P_T}),
avoiding numerically solving the FP equation (Eq.~\ref{eq:fp}).

Since the CRE is implemented in Lagrangian code Gadget-3, it is convenient to
normalise the physical quantities to mass instead of volume. 
Therefore we define
\begin{subequations} \label{eq:norm}
\begin{equation}
        \widetilde{C} = \frac{C m_e}{\rho},
\end{equation}
\begin{equation}
        \tilde{n} = \frac{n m_e}{\rho},
\end{equation}
\begin{equation}
        \tilde{\epsilon} = \frac{\epsilon }{\rho},
\end{equation}
\begin{equation}
        \widetilde{P} = \frac{P}{\rho},
\end{equation}
\begin{equation}
        \overline{T} = \frac{\tilde{\epsilon}}{\tilde{n}} m_e,
\end{equation}
\end{subequations}
where $\rho$ is the baryon density.

\subsubsection{Diffusive shock injection}\label{sec:inj}

\subsubsection*{a. Detecting shock waves}

    \mbox{\citet{Pfrommer_2006}} developed a formalism for the identification and accurate estimation of the strength of structure formation shocks on the fly in cosmological SPH-simulation. 
    As they pointed out, the grid-based techniques offer
    superior capabilities in capturing shocks, 
    while the dependence on the artificial viscosity
    is one drawback of SPH.
    Due to the broadening of shocks over the SPH smoothing
    scale, it can not be resolved as discontinuities,
    but the post-shock quantities can be
    calculated very accurately.
    We review their method of detecting shock
    in the following:

    The shock surface separates two regions: the upstream region and downstream region, from which physical quantities
    (such as density $\rho$ and pressure $P$) are labelled by 1 and 2, respectively.
    For a non-radiative polytropic gas, the conservation of mass, momentum, and energy flux allow us to
    derive the well-known Rankine-Hugoniot conditions
    \mbox{\citep[][]{Landau_1959, Pfrommer_2006}}:

\begin{subequations}
\begin{equation}
        \frac{\rho_2}{\rho_1} = \frac{\left(\gamma_a+1 \right) M_1^2}{\left(\gamma_a-1\right)M_1^2 + 2},
    \label{eq:RH1}
\end{equation}
\begin{equation}
        \frac{P_2}{P_1} = \frac{2 \gamma_a M_1^2 - \left(\gamma_a - 1\right)}{\gamma_a + 1},
    \label{eq:RH2}
\end{equation}
\begin{equation}
        \frac{T_2}{T_1} = \frac{\left[2 \gamma_a M_1^2\left(\gamma_a-1\right)\right]
        \left[\left(\gamma_a-1\right) M_1^2 + 2\right]} {\left(\gamma+1\right)^2 M_1^2},
    \label{eq:RH3}
\end{equation}
\label{eq:RH}
\end{subequations}
where
$T$ is the temperature, $M_1=v_1/c_{s1}$
is the Mach number in the upstream region
with $c_{s1} = \sqrt{\gamma_a P_1/\rho_1}$
being the speed of sound, and $\gamma_a$ being the adiabatic index.

Suppose that the shock is broadened to be the same order as the SPH smoothing length $f_h h$, where $f_h \sim 2$ is
a calibrated factor \mbox{\citep[see][]{Pfrommer_2006}} . The time for a particle to pass through the broadened
shock front is estimated as $\Delta t \approx f_h h / v_1$.
In Gadget, the entropic function is defined by $A \equiv P/{\rho^{\gamma_a}}$ \mbox{\citep{Gadget2}}.
The jump of the entropic function of particle
between the shock surface is estimated as 
\mbox{\citep[][]{Pfrommer_2006}}
\begin{equation}
    \label{eq:A2_A1_1}
        \frac{A_2}{A_1} = \frac{A_1 + \Delta t dA_1/dt}{A_1} = 1 + \frac{f_h h}{M_1 c_1 A_1} \frac{dA_1}{dt}.
\end{equation}
By substituting Eqs.~\mbox{\ref{eq:RH1}} and \mbox{\ref{eq:RH2}},
into Eq.~\mbox{\ref{eq:A2_A1_1}}, we have
\begin{equation}
    \label{eq:A2_A1_2}
        \begin{split}
            \frac{A_2}{A_1} &= \frac{P_2}{P_1} \left(\frac{\rho_1}{\rho_2}\right) ^ {\gamma_a} \\
            &= \frac{2 \gamma_a M_1^2 - \left(\gamma_a-1\right)}{\gamma_a + 1}
            \left[\frac{\left(\gamma_a-1\right) M_1^2 + 2}{\left(\gamma_a + 1\right)M_1^2}\right]^{\gamma_a}.
    \end{split}
\end{equation}
    By combining Eqs.~\mbox{\ref{eq:A2_A1_1}} and \mbox{\ref{eq:A2_A1_2}}, the final equation for estimating Mach number is
\begin{subequations}
\begin{equation}\label{eq:FA}
        \left[f_A(M_1) - 1\right] M_1 = \frac{f_h h}{c_1 A_1} \frac{dA_1}{dt},
\end{equation}
\begin{equation}
          f_A{M_1}  = \frac{2 \gamma_a M_1^2 - \left(\gamma_a-1\right)}{\gamma_a + 1}
            \left[\frac{\left(\gamma_a-1\right) M_1^2 + 2}{\left(\gamma_a + 1\right)M_1^2}\right]^{\gamma_a}.
\end{equation}
\end{subequations}
The right-hand side of Eq.~\ref{eq:FA} can be estimated individually for each particle, and the left-hand side depends only on $M_1$.  

For a composite of CRP and thermal gas,
the Mach number is derived with 
a similar procedure as the polytropic gas
(see section 3.2 of \mbox{\citet{Pfrommer_2006}} for more details).
In cosmological simulation, the Mach number statistics
generated by this method
agree well with the results obtained with hydrodynamics
mesh codes that use explicit Riemann solvers
\mbox{\citep[][]{Pfrommer_2006}}. In addition, 
this scheme has a good convergence with different resolutions
\citep{Pfrommer_2006,Vazza_2011}.

\subsubsection*{b. Injection}
\label{sec:inj2}
In DSA, particles are accelerated by multiple shock crossings
\citep{Fermi_1949}. The energy spectrum of suprathermal
electrons produced by DSA is well characterised by a power-law distribution. The
spectral index $\alpha_{\rm inj}$
is determined by the compression ratio at shock front,
i.e.
\begin{equation}
        \alpha_{\rm inj} = \frac{r+2}{r-1},
        \label{eq:r}
\end{equation}
where $r = \rho_2/\rho_1$ denotes the shock compression ratio with
$\rho_2$ and $\rho_1$ being the baryon density in downstream and upstream regimes of the shock, respectively.

We define the energy injection efficiency $\zeta_{\rm DSA}$ to be the energy density
ratio of freshly injected CRE to the total dissipated energy in the downstream regime,
\begin{subequations}
\begin{equation}
\label{eq:zeta}
        \zeta_{\rm DSA} = \frac{\epsilon_{\rm inj}}{\epsilon_{\rm dis}},
\end{equation}
\begin{equation}
        \epsilon_{\rm dis} = \epsilon_2 - \epsilon_1 r^{\gamma_a},
\end{equation}
\end{subequations}
where $\epsilon_1$ and $\epsilon_2$ are the energy density in
upstream and downstream region
of the shock, respectively, $\epsilon_{\rm inj}$ is the injected energy density,
and $\epsilon_{\rm dis}$ is
the dissipated energy density which is the
difference of the energy densities in
the pre-shock and pos-shock region.
In this work, we adopt $\zeta_{\rm DSA}=0.005$~\citep{Hoeft_2008}.

Even though we can account for CRE injection by shocks in SPH
using the Mach finder developed by \mbox{\citet{Pfrommer_2006}},
the shock broadening inherent in SPH is a problem,
to receive the full dissipative energy,
an SPH particle may require several timesteps before it has 
passed through a shock. 
\mbox{\citet{Jubelgas_2008}} have faced the same problem in their
DSA injection of CRP physics.
As they pointed out,
because the correct pre-shock and post-shock state fulfill the 
the conservation of energy in SPH code,
the correct integration of $\epsilon_{\rm dis}$
through the shock profile will be accomplished by SPH code
automatically.  Therefore
we can replace $\epsilon_{\rm dis}$
in Eq.~\ref{eq:zeta} with the dissipated energy
in the current timestep.
Note that we remove the injection energy $\epsilon_{\rm inj}$ from the thermal pool.

The minimum momentum $p_{\rm inj}$ of DSA injection is an
important parameter in determining the electron spectrum, 
because a significant fraction of energy, pressure, and number density
are carried by the lower-energy part of the CRE spectrum.
Following ~\citet{Hoeft_Bruggen_2007},
we adopt $p_{\rm inj} = 10 \, {k_{B} T}/{m_ec^2}$,
suggesting that $p_{\rm inj}$ is tightly coupled with 
the temperature of the plasma.

Since the DSA process is very efficient, after DSA injection,
we assume the lower cut-off, upper cut-off, and spectral index
to be $p_{\rm inj}$, $\infty$ and $\alpha_{\rm inj}$, respectively.
Thus the normalisation $\widetilde{C}$ is determined by numerically
solving the equation:
\begin{equation}
        \tilde{\epsilon}\left(\widetilde{C}, \alpha_{\rm inj},
        p_{\rm inj}\right) =
        \tilde{\epsilon}_{\rm inj}+\tilde{\epsilon}_{\rm old}.
\end{equation}
Note that we suppose that the injected spectrum
and the new spectrum have no upper cut-off.
Since we use the conservation of energy to derive the spectrum
parameters in the injection process,
the results of injection do not depend on the upper cut-off.

\subsubsection{Loss}\label{sec:cool}
    In this subsection, we analyse the cooling processes
    (Eqs. 
    \mbox{\ref{eq:dpdt_rad}},
    \mbox{\ref{eq:dpdt_coul}}
    ) and describe the methods to determine the
    upper cut-off $p_{\max}$ and lower cut-off $p_{\min}$.

\subsubsection*{a. The upper cut-off $p_{\max}$}

    \citet{Lawson_1987} has suggested that
    the upper energy limit of the DSA account for the steepening of the radio spectrum (Figure 7 of \citet{Lawson_1987}). 
    In this work, we assume that there is no upper cut-off for
    DSA injection (see Sec.~\ref{sec:inj2}) and attribute 
    upper cut-off to the radiative losses
    (Eq.~\mbox{\ref{eq:dpdt_rad}}),
    which dominate at high energy regime.
    Considering that the momentum of an electron decreases from $p_0$
    at time $t_0$ to $p_1$ at time $t_1$, the conservation of energy
    gives:

\begin{equation}
        \int_{p_0}^{p_1}\frac{dp}{p\sqrt{1+p^2}} = 
        -\int_{t_0}^{t_1}  C_{\rm cool}(t) dt,
    \label{eq:single_cool_rad}
\end{equation}
where the time dependence of $C_{\rm cool}$ comes from the baryon magnetic field \mbox{\citep{Dolag_Stasyszyn_2009}}
and the the equivalent magnetic field of CMB.
If the initial spectrum is a power law, then the final spectrum
has a maximum momentum $p_{\max}$ given by:

\begin{equation}
        \int_{\infty}^{p_{\max}}\frac{dp}{p\sqrt{1+p^2}} = 
        -\int_{t_0}^{t_1}  C_{\rm cool}(t) dt,
\end{equation}
that is
\begin{equation}
        \frac{1}{p_{\max}} = {\rm sinh}\left( \int_{t_0}^{t_1}  C_{\rm cool}(t) dt\right),
\end{equation}
where the the right hand side is computed by the accumulation of $C_{\rm cool}(t)$ in simulation.
We use this maximum momentum $p_{\max}$ as our upper cut-off.

\begin{figure}
    \includegraphics[width=\myfigwidtha]{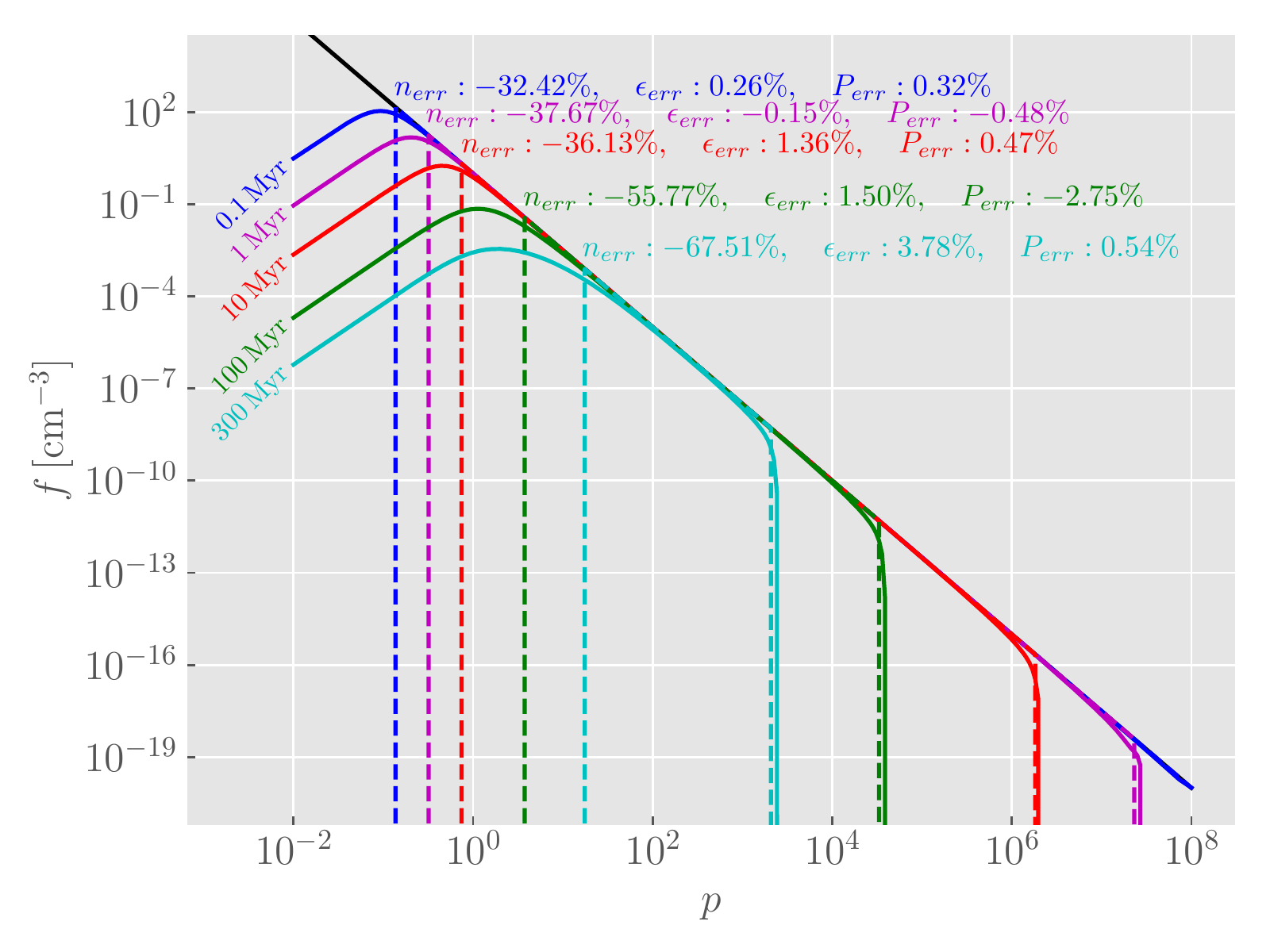}
    \caption{\
        Evolution of CRE distribution. The initial CRE populations are discribed by
        $(C, \alpha, p_{\min}, p_{\max}) = (1, 2.5, 10^{-2}, 10^8)$. Spectra are shown for cooling
        ages of $\approx (0.1, 1, 10, 100, 300)\,{\rm Myr}$. The solid lines show the numerically exact solutions
        and the model solutions are displayed by dashed lines. 
        $n_{\rm err}, \epsilon_{\rm err}$ and $P_{\rm err}$ are the relative difference of
        number density, energy density and pressure of CRE, respectively.
    }
    \label{fig:comp}
\end{figure}

\begin{figure*}
    \includegraphics[width=\myfigwidthb]{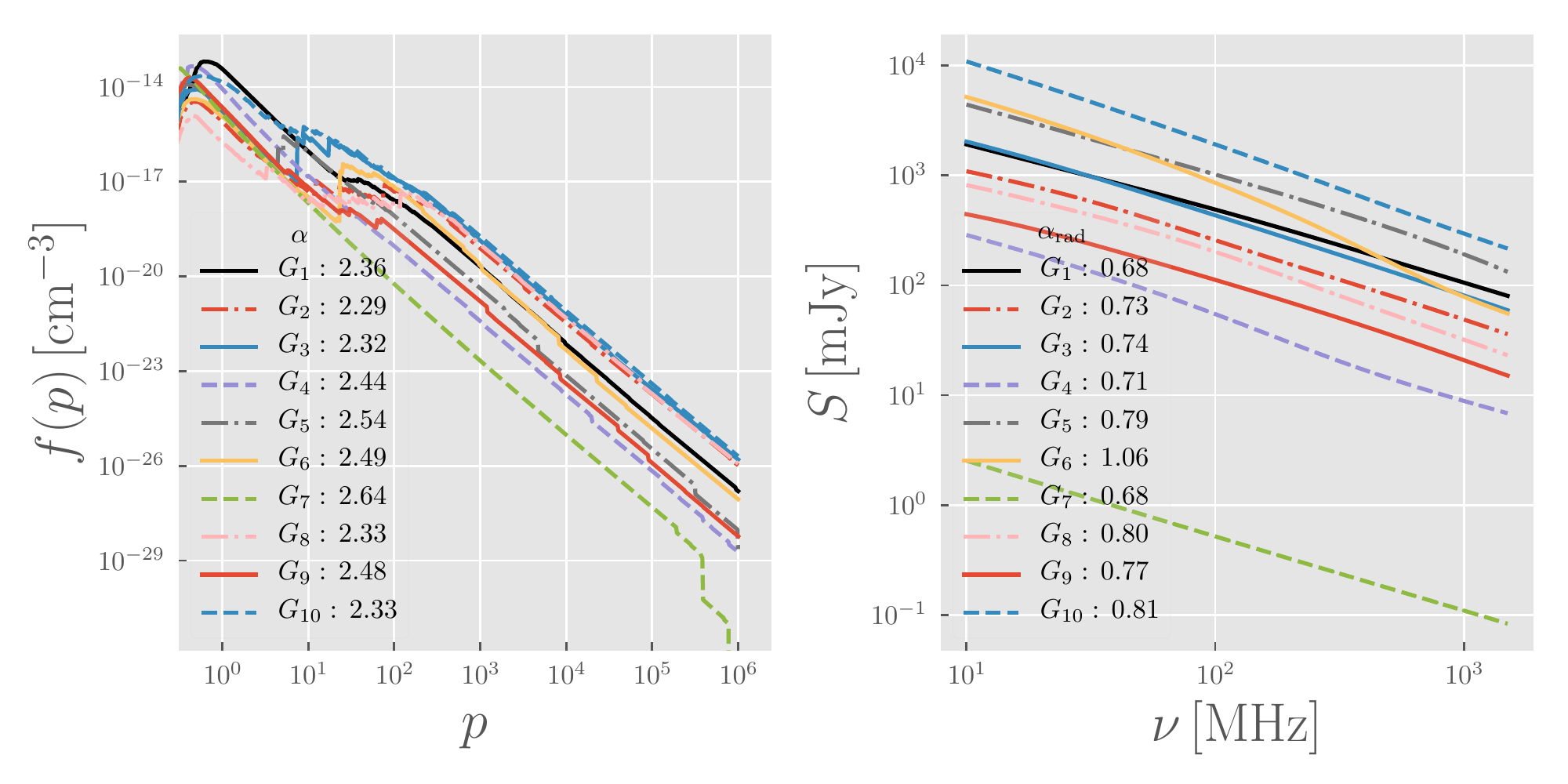}
    \caption{\
    Integrated CRE distribution (left panel) and radio emission spectrum(right panel) of ten most massive 
    clusters with $M > 10^{14} M_{\odot}$
    at $z=0.1$.
    ~$\alpha$ is the slope of electron spectrum in the momentum range $10^3-10^5$.
    ~$\alpha_{\rm rad}$ is the spectral index of radio emission in
    the frequency range $\rm 0.3 - 1.4 \, GHz$.
    }
    \label{fig:ele_rad_spec}
\end{figure*}
\begin{figure*}
    \includegraphics[width=\myfigwidthb]{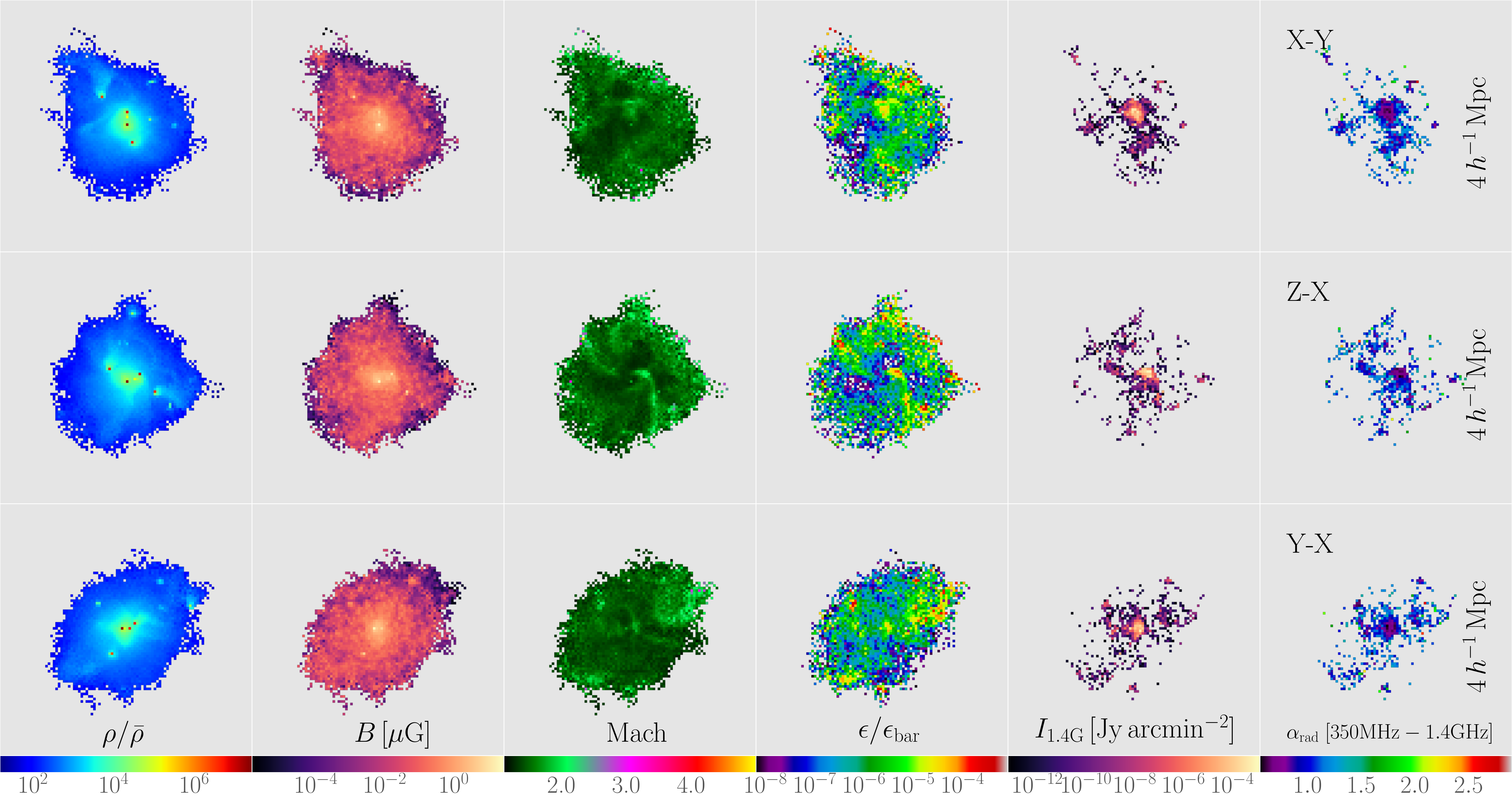}
    \caption{\
    The three direction projection for 
    the most massive cluster in SIM-CRE simulation
    taken from snapshot with $z=0.1$.
    From left to right,
    baryon density,
    magnetic field,
    Mach number,
    CRE energy,
    radio emission  of $\rm 1.4GHz$,
    and the spectral index in $\rm 350 MHz - 1.4GHz$
    are shown.
    The projection cube has a comoving side length
    $5\, h^{-1} \rm Mpc$.
    $\epsilon_{\rm bar}$ is the baryon energy.
    }
    \label{fig:group}
\end{figure*}

\begin{figure}
    \includegraphics[width=\myfigwidtha]{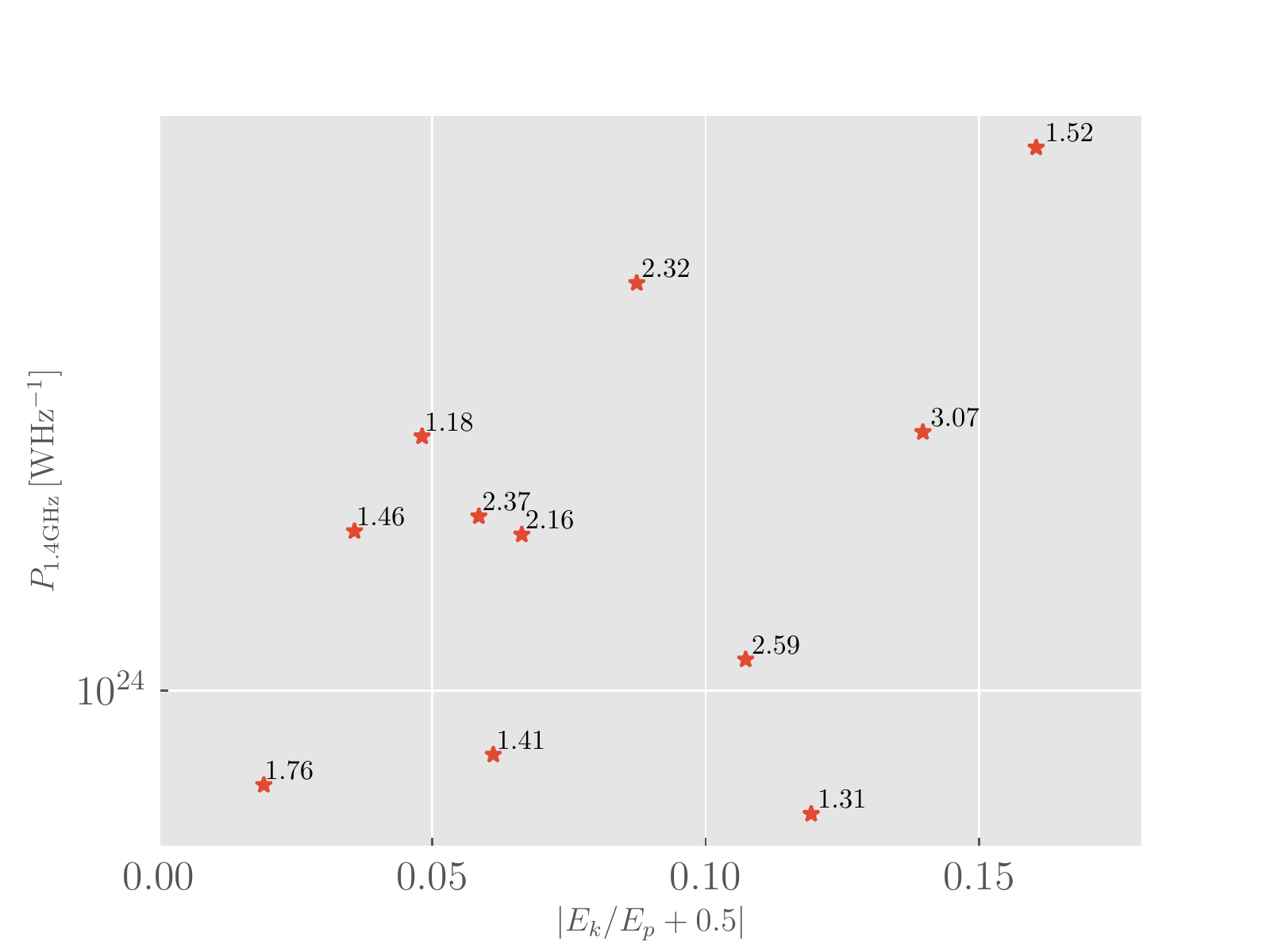}
    \label{fig:rad_vir}
    \caption{\
    Ralation between radio power $P_{\rm 1.4 GHz}$ and
    the virialization state for some massive clusters with
    $P_{1.4 \, \rm GHz} > 5\times 10^{23} \rm W Hz^{-1}$ 
    and mass $M > 10^{14} h^{-1} M_\odot$,
    where
    $E_k$ and $E_p$ 
    are the kinetic energy and potential energy.
    }
    \label{fig:rad_vir}
\end{figure}

\begin{figure}
    \includegraphics[width=\myfigwidtha]{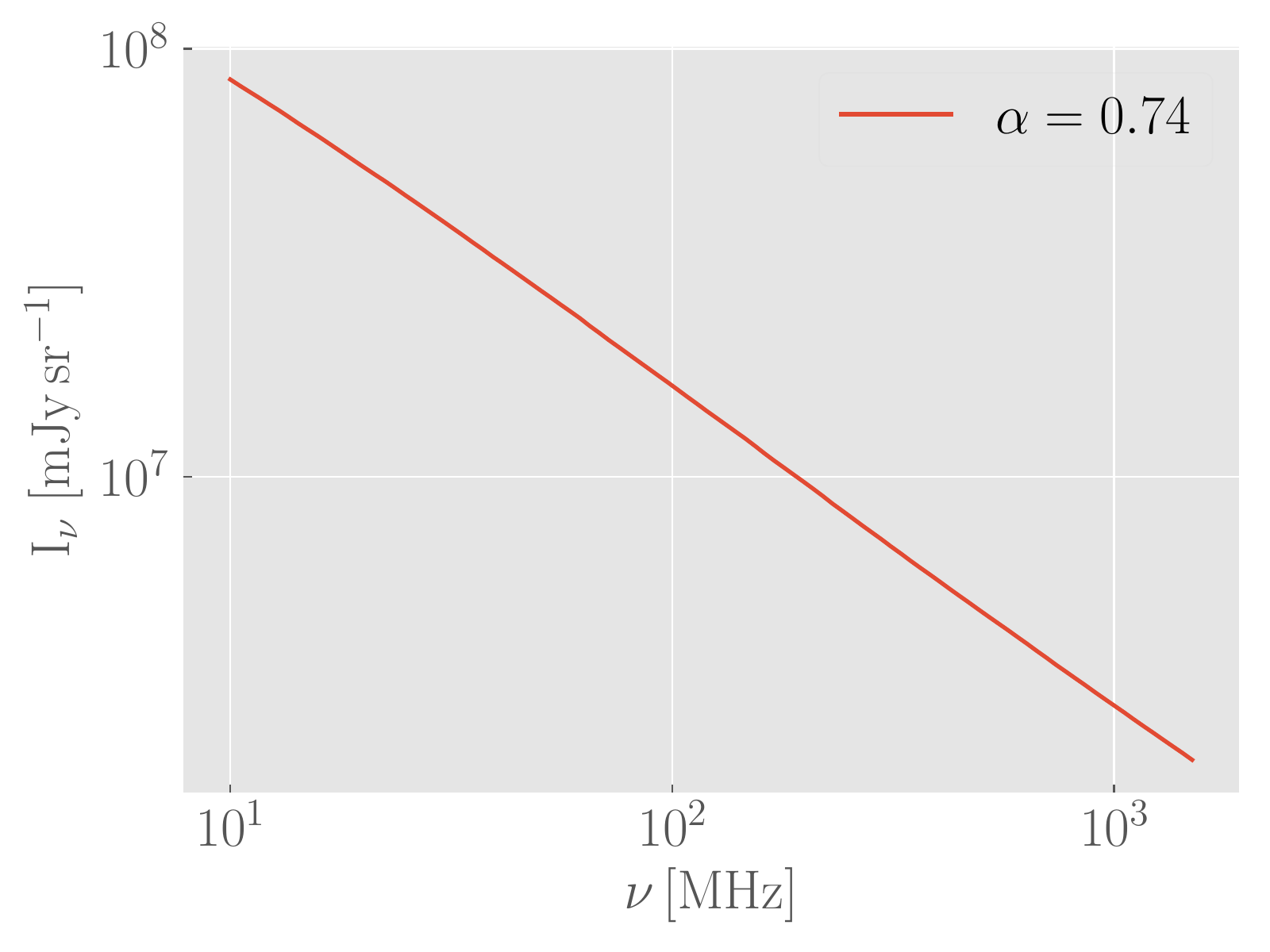}
        \caption{The radio background emission obtained from our
         SIM-CRE simulation.}
    \label{fig:totspec}
\end{figure}

\begin{figure}
    \includegraphics[width=\myfigwidtha]{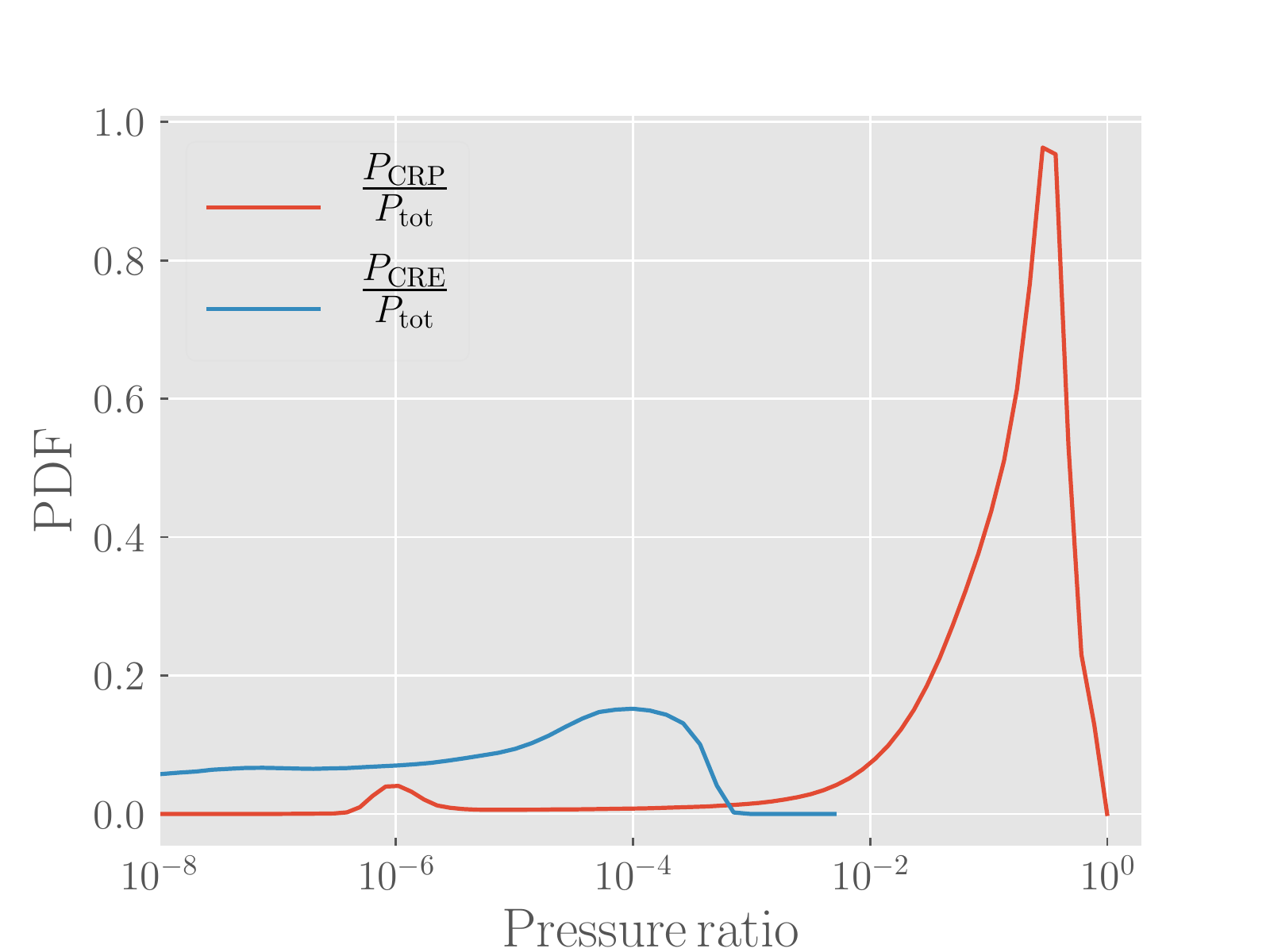}
        \caption{\
        Probability Density Function (PDF)
        of the CRE pressure and CRP pressure of
        particles taken from the snapshot with reshift $z=0$.
        }
    \label{fig:ppdf}
\end{figure}

\begin{figure*}
    \includegraphics[width=\myfigwidthb]{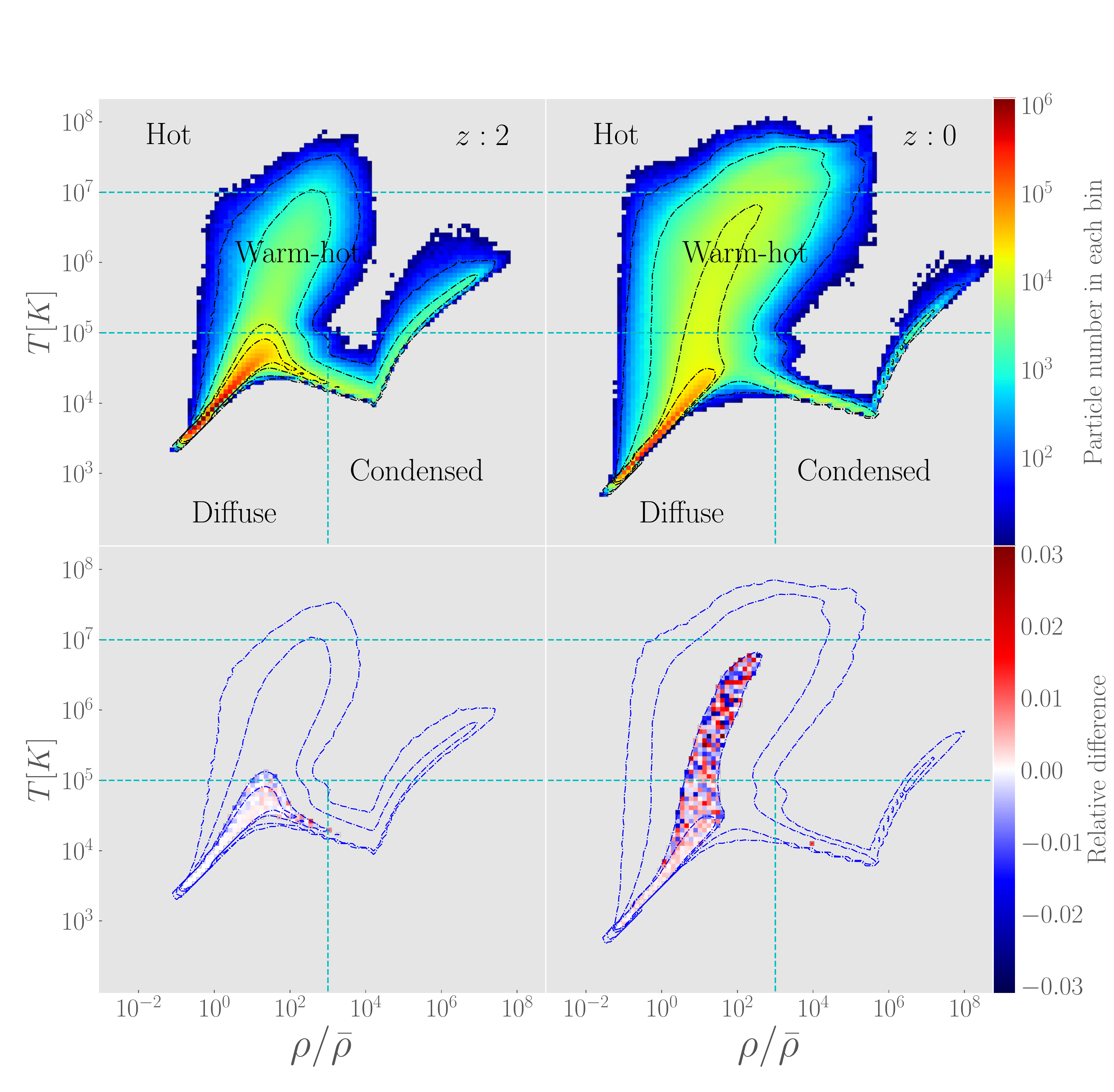}
        \caption{\
        Top: Gas density-temperature phase diagram of SIM
        simulation at $z=2$ (left) and $z=0$ (right). 
        $\bar{\rho}$ is the mean baryonic density,
        and Contour leves (dashdot line) are placed at $10^2, 10^3, 10^4$ and $2\times 10^4$.
        Bottom: Relative difference of gas phase diagram between
        SIM and SIM-CRE.
        In order to avoid the statistical error,
        we only plot the relative difference
        for the bins with the particle number
        larger than $10^4$,
    \textit{Diffuse}:
    ${\rho}/\bar{\rho} < 1000, T < 10^5 {\rm K}$.
    Photoionized intergalactic gas.
    \textit{Condensed}:
    ${\rho}/\bar{\rho} > 1000, T < 10^5{\rm K}$.
    Stars and cool galactic gas.
    \textit{Warm-hot}:
    $ 10^5 {\rm K} < T < 10^7 {\rm K}$.
    Warm-hot intergalactic medium.
    \textit{Hot}: $T > 10^7 {\rm K}$.
    Gas in galaxy clusters and large groups
    \citep{Dave_2001}.
}
    \label{fig:gaspha}
\end{figure*}

\subsubsection*{b. The lower cut-off $p_{\min}$}

Since most energy and pressure are carried by low-momentum electrons
and the energy and pressure are two ways for simulation to interact with
the CRE processes, the lower cut-off $p_{\min}$ of our model must guarantee the accurate calculations of energy and
pressure. On the other hand, the normalisation of spectrum
is very sensitive to the lower cut-off for fixed total energy,
so the lower cut-off will affect the radio emission.
At low energy, the dominant losses will be the Coulomb loss.
The calculation of Coulomb loss (Eq.\mbox{\ref{eq:dpdt_coul}}) is complicated
and some approximation should be adopted.
To this end, we replace the momentum $p$ in the curly braces of
Eq.\mbox{\ref{eq:dpdt_coul}} with 
its mean value for the given initial spectrum being 
$\langle p \rangle = p_{\min,0}(1-\alpha)/(2-\alpha)$,
where $p_{\min,0}$ is the lower cut-off of the initial spectrum.
Then, the dependence of momentum in the right hand side of Eq.~\ref{eq:dpdt_coul} is only in $\beta$ and  Coulomb loss function,
thus Eq.\mbox{\ref{eq:dpdt_coul}} becomes

\begin{equation}
        \frac{dp}{dt}\bigg|_{\rm coul} \approx -\frac{C_{\rm coul}
        \left(\langle p \rangle, t\right)}{\beta^2},
    \label{eq:dpdt_coul_approx}
\end{equation}
    where  $C_{\rm coul}$ is given by Eq.~\mbox{\ref{eq:dpdt_coul}} and 
    does not depend on the momentum $p$. The time dependence of $C_{\rm coul}$ comes
    from the number density of thermal electron $n_{\rm th, e}$. 
    Considering that the momentum of an electron decreases from $p_0$
    at time $t_0$ to $p_1$ at time $t_1$, the conservation of energy
    gives:
\begin{equation}
        \int_{p_0}^{p_1} \beta^2 dp \approx -\int_{t_0}^{t_1}
        C_{\rm coul}\left(\langle p \rangle, t\right) dt.
\end{equation}

If the electrons with momentum $p<p_{\rm cut}$ in 
an initial spectrum  at time $t_0$ do not appear in
the final spectrum at time $t_1$,
then $p_{\rm cut}$ is given by
\begin{equation}
        \int_{p_{\rm cut}}^{0} \beta^2 dp 
        = p_{\rm cut} - {\rm arctan}(p_{\rm cut})
        \approx -\int_{t_0}^{t_1}
        C_{\rm coul}\left(\langle p \rangle, t\right) dt,
\end{equation}
where the integral on the left hand side is computed 
by the accumulation of $C_{\rm coul}(\langle p \rangle, t)$ in simulation.

    Since the Coulomb losses vary slowly with $p$ and
    result in a flat spectrum (see Fig.\mbox{\ref{fig:comp}})
    at the low energy, using the $p_{\rm cut}$ as the lower
    cut-off will lead to an underestimate of the energy loss.
    Therefore we should consider the energy loss
    of electrons with momentum $p \ge p_{\rm cut}$.
    From Eq. \mbox{\ref{eq:dpdt_coul_approx}}
    we can obtain the energy loss rate for a single electron:
\begin{equation}
        \frac{dT(p)}{dt}\bigg|_{\rm coul} = m_ec^2 \beta \frac{dp}{dt}\bigg|_{\rm coul}
        \approx -m_ec^2 C_{\rm coul}(\langle p \rangle,t)\frac{1}{\beta}.
\end{equation}
The energy losses of a single electron from time $t_0$ to $t_1$ is given by
\begin{equation}
        \Delta T(p) = \int_{T_0}^{T_1} dT(p) dp \approx 
        -\frac{1}{\beta} \int_{t_0}^{t_1} m_ec^2 C_{\rm coul}(\langle p \rangle,t) dt,
\end{equation}
where $T_0$ and $T_1$ are the energy of the single electron at time $t_0$ and $t_1$, respectively.
In order to obtain the total Coulomb loss of the CRE spectrum, we have to integrate
above equation over the population $f(p)$, i.e.
\begin{equation}
        \begin{split}
            \Delta  \tilde{\epsilon} &= \frac{\Delta \epsilon}{\rho} \approx
            -\frac{1}{\rho}\int_{p_{\rm cut}}^{\infty} \frac{f(p)}{\beta}dp
            \int_{t_0}^{t} m_ec^2 C_{\rm coul}(\langle p \rangle, t) dt
            \\
            &=-\frac{\widetilde{C}}{\alpha}\left[ \frac{1}{2}
            B_{\frac{1}{1+q_{\rm cut}}}\left(\frac{\alpha-1}{2}, \frac{2-\alpha}{2}\right)
            + q^{-\alpha}
            \sqrt{1+q^2} \right]
            \\
            &\times \int_{t_0}^{t} C_{\rm coul}(\langle p \rangle, t) dt.
        \end{split}
\end{equation}
Consequently, the lower cut-off of our model is determined by
numerically solving the  following equation:
\begin{equation}
        \tilde{\epsilon}\left(\widetilde{C}, \alpha, p_{\min}\right) =
        \tilde{\epsilon}\left(\widetilde{C}, \alpha, p_{\rm cut}\right) +
        \Delta \tilde{\epsilon}\left(\widetilde{C}, \alpha, p_{\rm cut}\right).
\end{equation}
    Opposite to the radiative loss, this cooling energy
    will be returned to the thermal pool. Note that we assume 
    that $\widetilde{C}$ and $\alpha$ remain unchanged
    (see Fig.~\ref{fig:comp}).

\subsubsection*{c. Approximation validation}\label{sec:acctest}
To validate the accuracy of our model, we 
solve the FP equation numerically for a freely
cooling via a Crank-Nicholson scheme with a constant
timestep of $\Delta t = 10^{-4}\, \rm Myr$
and 300 logarithmic momentum points. The thermal electron
number density varies linearly 
from $10^{-4}$ to $10^{-3}\,\rm cm^{-3}$ and
the magnetic field increase linearly
from $1$ to $10\, \rm \mu G$.
We adopt a time interval of $\rm 0.1 \, Myr$ to compute
the cut-offs, which approximate the 
timestep in hydro-simulation. From the analysis to determine
the cut-offs,
unlike the finite difference method, which is known to be
stable for a certain timestep, 
our scheme of treating CRE is insensitive to the adopted
timestep and suitable for the usage in hydro-simulation.
We show the numerically exact solution (solid lines) and
approximate (dashed lines) distribution with an initial
population described by $(C, \alpha, p_{\min}, p_{\max})
= (1, 2.5, 10^{-2}, 10^8)$ in Fig.\mbox{\ref{fig:comp}},
from which we find that the approximate
treatment captures the evolution of the exact solution
reasonably well, 
the errors of energy and pressure are $\lesssim 4\%$ within
$300 \rm Myr$.
The discrepancy at low energy is acceptable for the present
purpose since those electrons hardly contribute to the
radio emission and the hydro-simulation only cares about
the energy and pressure of CRE.
Because the low energy part of the electron population
is ignored, it is not surprising that
the total CRE number density is underestimated.
For some applications, the accuracy level of our
approximation method should be
sufficient \citep{Ensslin_2007},
even though the energy difference seems to increase
after a larger timescale.
A more sophisticated treatment of CRE physics may be needed for 
some application with a requirement of high accuracy level,
which is beyond the scope of present work.

\subsubsection{Adiabatic energy changes}\label{sec:adia}
If the electrons are confined within a varying volume, they are subject to adiabatic gains
and losses, which are described by
\begin{equation}
    \frac{df(p,t)}{dt} = -(\nabla \cdotp v) f(p,t).
\end{equation}
This processes has no effect on the cut-offs and spectral slope and leads to
\begin{equation}
    \frac{\widetilde{C}_{t+\Delta t}}{\widetilde{C}_t} =
    \frac{\tilde{n}_{t+\Delta t}}{\tilde{n}_t} =
    \frac{\tilde{\epsilon}_{t+\Delta t}}{\tilde{\epsilon}_t} =
    e^{-\nabla \cdotp v \Delta t}.
\end{equation}

\section{Simulation}\label{sec:simu}

\subsection{Simulation setup}
For our simulations, we adopt the same cosmological parameters as \citet{Marinacci_2015}:
$\Omega_{\rm m}=\Omega_{\rm dm}+\Omega_{\rm b}=0.302$,
$\Omega_{\rm b}=0.04751$,
$\Omega_{\Lambda}=0.698$,
$\sigma_8 = 0.817$,
$n = 1$
and $H_0=\rm 68\, km\,s^{-1}\,Mpc^{-1}$.

    \citet{Pfrommer_2006} have pointed out that their scheme of
    detecting shock has good convergence and
    used a simulation employed $2\times 256^3$
    particles in a periodic box of comoving size
    $100 \, h^{-1} \, \rm Mpc$ to study the cosmological shock waves.
    \mbox{\citet{Jubelgas_2008}} have used
    this scheme and picked a
    comoving box of side-length $100\, h^{-1} \rm Mpc$
    to simulate their CRP model at two resolutions,
    with $\rm 2 \times 128^3$ and $\rm 2 \times 256^3$
    particles, respectively. 
    As they pointed out that the results of their two resolutions
    are in good agreement with each other.
    Since the injection source of their CRP model is also the DSA
    and we use the same method to detect shock
    waves as them,
    we chose the resolution with a comoving box of side-length
    $100\, h^{-1} \rm Mpc$ and $2 \times 256^3$ particles.

    We run two cosmological MHD simulations, named as
    SIM (without CRE physics) and SIM-CRE (with CRE physics,
    Fig.~\ref{fig:slices} of appendix~\mbox{\ref{sec:visual}}
    gives some visualization).
    Initial condition with $z=127$
    is created by the code 2LPTIC \mbox{\citep{2lpt_2006}}  with
    an Efstathiou power spectrum \mbox{\citep{Efstathiou_1992}},
    which is based on second-order Lagrangian Perturbation Theory (2LPT),
    rather than first-order (Zel'dovich approximation). 
    In order to compute the radio background from intergalactic shocks
    (see Sec.~\mbox{\ref{sec:rad_obs}}), we output 142 snapshots
    within the range of redshift $15-0$.

    We run Gadget-3 with the default setting of the numerical SPH parameters,
    using 32 neighbours in smoothed estimates and an artificial viscosity 
    parameter of $\alpha =0.8$, combined with Balsara's switch \mbox{\citep{Balsara_1995}}
    to reduce the viscosity in the presence of strong shear.
    The baryon physics included in our simulation are
    star formation, cooling processes, shock wave, CRP, and MHD,
    the settings of which are:
    \textbf{(1) Star formation}, we adopt the model parameters suggested by
    \mbox{\citet{Springel_Hernquist_2003}}
            and take the number of stars each gas particle may form
            as 1 \mbox{\citep[see][]{Springel_Hernquist_2003}}.
    \textbf{(2) Cooling}, we use the default cooling scheme, the cooling rates of which
            are given by \citet{Katz_1996}.
    \textbf{(3) Shock waves}, we use a composite of CRP and thermal gas to derive Mach number and 
            take shock length scale parameter $f_h$ as 2 \citep{Pfrommer_2006}.
    \textbf{(4) CRP}, we take the parameters advised by \citet{Jubelgas_2008} for the CRP spectrum,
            the injectons of DSA and supernovae.
    \textbf{(5) MHD}, we use the MHD implementation of \mbox{\citet{Dolag_Stasyszyn_2009}}
            with the hyperbolic/parabolic divergence cleaning scheme
            originally proposed by \mbox{\citet{Dedner_2002}},
            which has found popular use in in both Eulerian \mbox{\citep{Mignone_2010}}
            and Lagrangian codes \mbox{\citep{Pakmor_2011}},
            to ensure the $\nabla \cdot B = 0$ constraint,
            and a limiter proposed by \mbox{\citet{Stasyszyn_2013}}
            to avoid overcorrections due to the cleaning scheme.
            The hyperbolic, parabolic and limiter paramter are set to $4$, $2$ and $0.5$,
            respectively \mbox{\citep{Stasyszyn_2013}}.
            In the case of adiabatic the magnetic field evolves as
            $B=B_0(1+z)^2 \propto \rho^{2/3}$, where $B_0$
            is the rescaled intensity of the $B$ at $z=0$ or the comoving magnetic field, $\rho$
            is the gas density, the structure formation will amplify $10^{-14}$ comoving Gauss
            seed fields to the value observed in low-reshift galaxies
            \mbox{\citep{Marinacci_2015, Marinacci_2018}}, so we use $10^{-10} \rm \, G$
            as our initial physical magnetic field at $z=127$.

\subsection{Radio emission and observation} \label{sec:rad_obs}

In order to verify the rationality of our model,
we discuss the computation of radio emission and several results
given by our SIM-CRE simulation.

The synchrotron power of a single electron with momentum $p$ in a magnetic field
$B$ is \citep[see][]{Rybicki_Lightman_1979,Hoeft_Bruggen_2007}
\begin{subequations} \label{eq:rad1}
\begin{equation}
        \frac{dP(p, \nu)}{d\nu} = \frac{\sqrt{3} \, B \, e^3 \, \sin \,  \alpha}{m_ec^2} F
        \left(\frac{\nu} {\nu_c}\right),
\end{equation}
\begin{equation}
        F(x) = x\int_{x}^{\infty}K_{5/3}(\xi) d\xi,
        \label{eq:F_x}
\end{equation}
\begin{equation}
        \nu_c = \frac{3 \, (1+p^2) \, e \, B \, \sin \, \alpha}{4 \pi m_e c},
\end{equation}
\end{subequations}
where $\alpha$ is the pitch angle,
$K_{{5}/{3}}$ is the modified Bessel function,
and $\nu_c$ is the characteristic frequency.
The synchrotron emissivity per volume is given by
\begin{equation}
    \frac{d^2 P(\nu)}{dV d\nu} = \int_{0}^{\infty} \, f( p ) \frac{dP(p,\nu)}{d\nu} dp.
    \label{eq:rad2}
\end{equation}

In SPH, we estimate the synchrotron power of 
an individual SPH particle at frequency $\nu$ by 
\begin{equation}
    P_{\rm sph}(\nu) = \frac{4}{3} \pi h^3 \frac{d^2 P(\nu)}{dV d\nu}
    \label{eq:syn-pow},
\end{equation}
where $h$ is the smoothing length of an SPH particle.

In Fig.~\mbox{\ref{fig:group}}, we plot  
baryon density, magnetic field, Mach number, CRE energy,
$\rm 1.4GHz$ radio emission
and the spectral index
in $\rm 350 MHz - 1.4GHz$
(from left to right)
of most massive cluster at $z=0.1$ in the
SIM-CRE simulation.
The magnetic field follows the baryon density
distribution, which reaches the largest value 
at the baryon density peak and
decreases quickly with baryon density.
There are also some local increases of
magnetic field outside the center of the cluster, which corresponds to
infalling sub-structures \citep{Marinacci_2015}.
The CRE energy is very related to
the shocks, this is because the only
injection source of CRE is shock in our model.
The ratio of CRE energy to baryon energy is
$\lesssim 0.1\%$, which is consistent with
the DSA injection efficiency $\zeta_{\rm DSA} = 0.005$.
Due to a weak magnetic field,
the radio emission of most CRE is very weak.
In Fig.~\ref{fig:rad_vir}, we demonstrate the relation between
radio power of $\rm 1.4 \, GHz$
and the virialization of cluster, where $E_k/E_p$ is the
virial ratio which is a direct measure of the 
dynamical state of a cluster, the radio power tends to 
increase with increasing $|E_k/E_p + 0.5|$, that is
relaxed clusters show much lower radio emission \citep{Buote_2001}.
The spectral index is a powerful tool to understand the physical properties of radio objects. 
The spectral data of halo and relic are reported in
Table 2 and Table 4 of \citet{Feretti_2012}, respectively.
From Table 2, Table 4 and Figure 18 of \citet{Feretti_2012} we 
know that radio objects have a spectral index
within a range $\sim 0.8-3$.
From last column of Fig.~\ref{fig:group},
we know that the spectral indices produced by our simulation
are consistent with that.

In Fig.~\ref{fig:ele_rad_spec},
we give the integrated CRE distribution and
the radio emission of ten most massive clusters.
The CRE spectrum is also a power law at high energy.
Since the low energy part of the CRE spectrum is ignored
in our model and the integrated CRE spectrum of a cluster
is obtained by a summation,
it is not surprising that
the flattening effect of Coulomb scattering in low energy
(Fig.~\mbox{\ref{fig:comp}}) can not be well described by our model,
however, this discrepancy does not affect the simulation
and the computation of radio emission (see Sec.\mbox{\ref{sec:acctest}}).
The spectral index of the integrated radio emission 
with frequency $\rm 100 MHz$ to a few $\rm GHz$
has been estimated to be in the range $0.7-0.8$
\citep[see Section 3.2.2 of][]{Keshet_2004}.
From Fig.~\ref{fig:ele_rad_spec}, 
most of the spectral index of
integrated radio emission produced by our SIM-CRE simulation are in
this range.

Finally, we estimate the radio background emission from our SIM-CRE simulation.
The radio background
from all particles in the simulation snapshot at redshift $z$ is 

\begin{equation}
        I(\nu_{\rm obs}, z) = 
        \frac{1}{4 \pi D_{\rm lum}^2(z) \Omega}
        \sum_{i=0}^{N_{\rm gas}}
        P_{\rm sph, i}[(1+z)\nu_{\rm obs}] 
    \label{eq:I_obs_z}
\end{equation}
where $D_{\rm lum}$ is the luminosity distance
and $\Omega$ is the solid angle of the simulation box.
To obtain the total radio background, one has to integrate
Eq.~\mbox{\ref{eq:I_obs_z}} from the low redshift $z_1$ to the high
reshift $z_2$, that is

\begin{equation}
\begin{split}
        I_{\rm tot}(\nu_{\rm obs}) &=
        \int_{z_1}^{z_2} 
        dI\left(\nu_{\rm obs},z\right)
        =
        \int_{z_1}^{z_2} 
        \frac{dI\left(\nu_{\rm obs}, z\right)} {dD_{\rm com}(z)}
        \frac{ dD_{\rm com}(z)}{dz} dz \\
        &\approx
        \int_{z_1}^{z_2} 
        \frac{I\left(\nu_{\rm obs}, z\right)} {L}
        \frac{ dD_{\rm com}(z)}{dz} dz
        \label{eq:tot_spec}
\end{split}
\end{equation}
where $D_{\rm com}$ is comoving distance.
In this work, we adopt $z_1=0.1$ and $z_2=5$.
Based on the dimensional-analysis model of \citet{waxman_2000},
\citet{Keshet_2004} have estimated the extragalatic radio
emission from the strong shocks involved in structure
formation. They have predicted that the radio 
in the frequency range $\rm 10-100MHz$ is in a 
range $\rm 10^8-10^7 mJy \, sr^{-1}$ with a
spectral index $\sim 1$ \citep[see Figure.6 of][]{Keshet_2004}.
We show the radio background emission
estimated from Eq.~\ref{eq:tot_spec} in
Fig.~\mbox{\ref{fig:totspec}},
from which we learn that the spectral
index of our result is $0.74$ and the intensity in
$\rm 10-100 MHz$ also ranges $\rm 10^8-10^7 mJy \, sr^{-1}$. 
Note that the intensity of radio background emission is
related to the DSA injection efficiency $\zeta_{\rm DSA}$,
so it can be regulated by $\zeta_{\rm DSA}$.

\begin{figure}
    \includegraphics[width=\myfigwidtha]{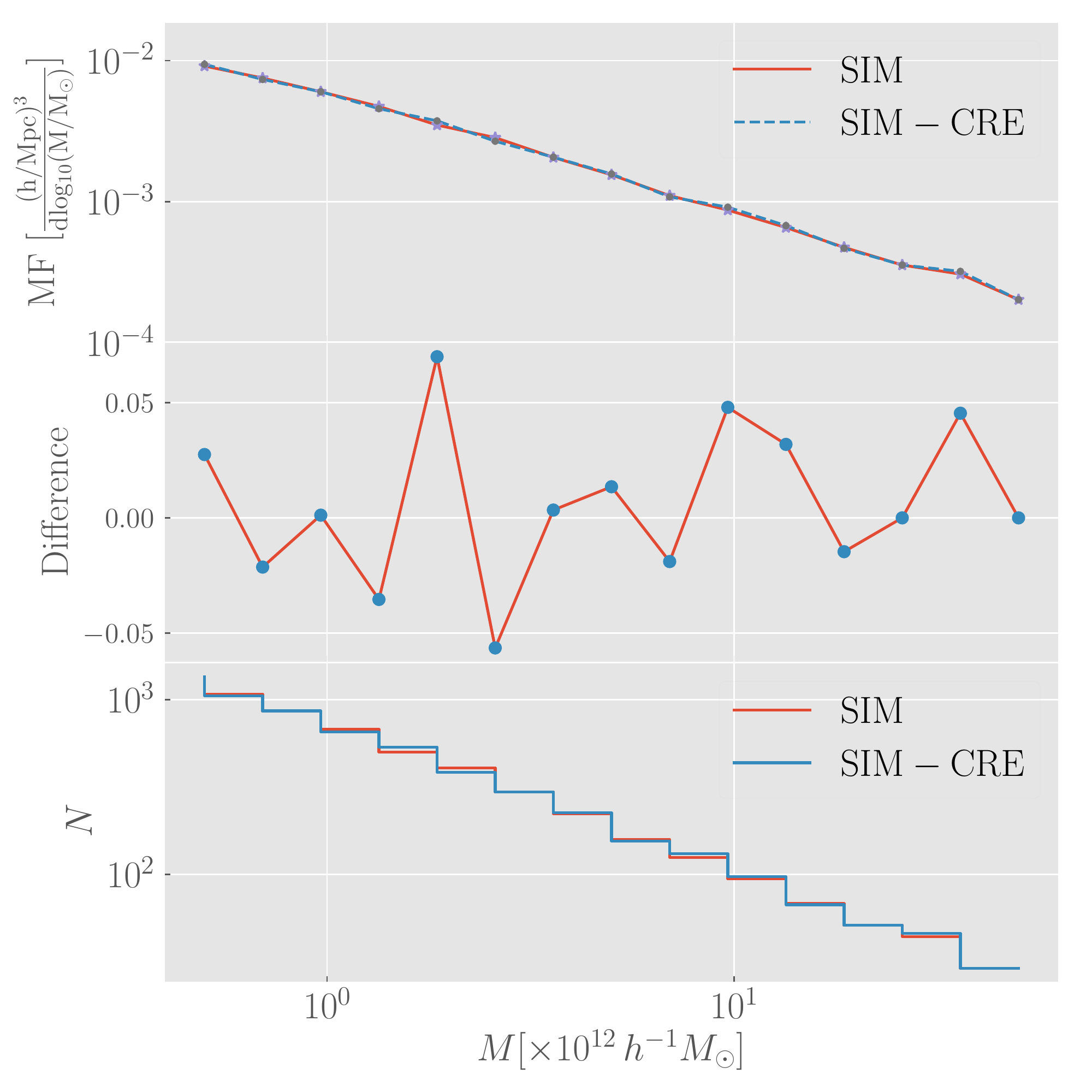}
        \caption{\
        Top: Mass function (MF) at z=0.
        Middle: Relative difference of MF between SIM and SIM-CRE.
        Bottom: the number of group in each bin.
        }
    \label{fig:mf}
\end{figure}

\begin{figure}
    \includegraphics[width=\myfigwidtha]{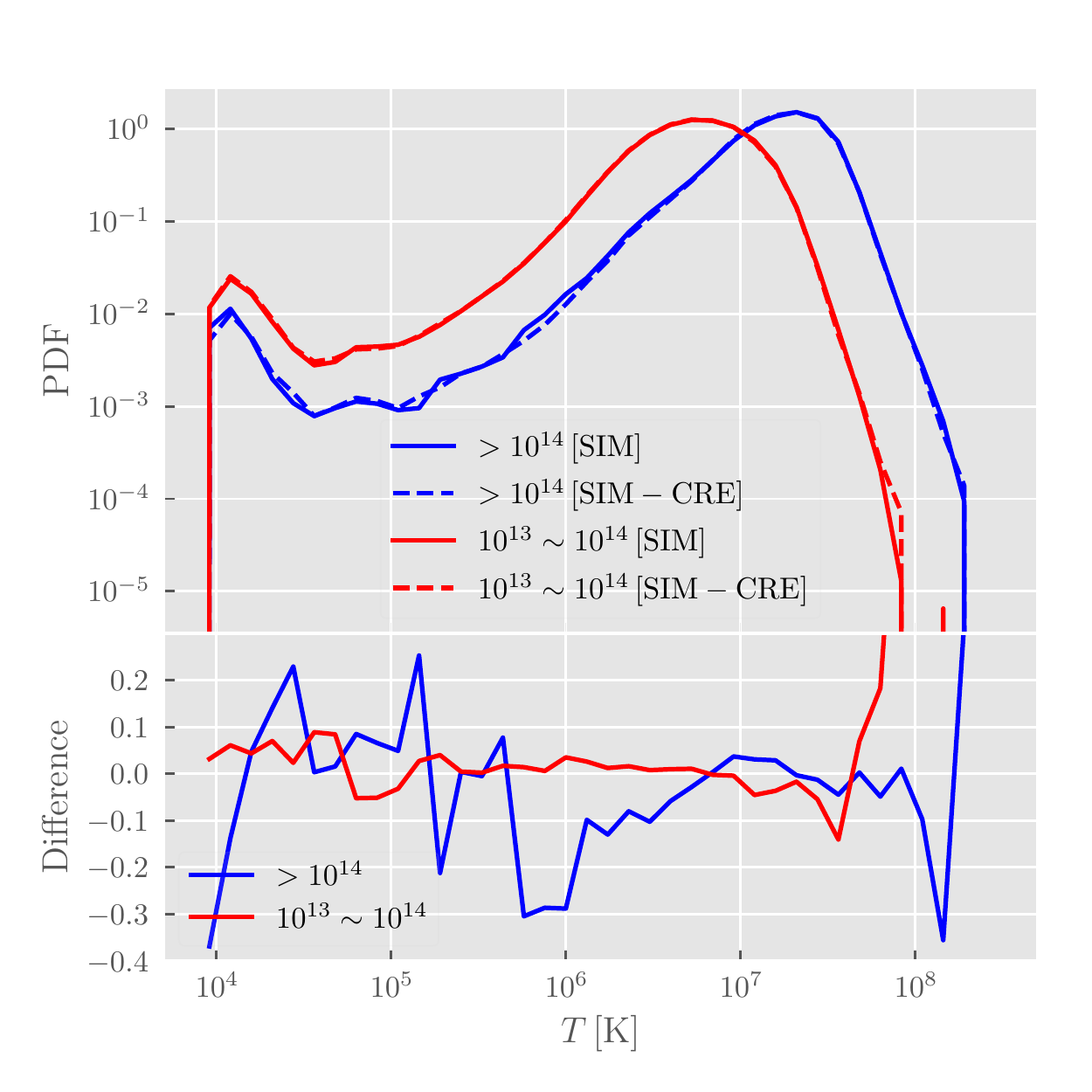}
        \caption{\
        PDF of the gas temperature in the FoF group
        at $z=0$.
        The blue lines in the top panel
        is the PDF of the gas temperature of all
        gas particles in the FoF groups with
        $M > 10^{14} h^{-1} M_\odot$, while
        the red lines is the slimilar PDF for
        the FoF groups with
        $10^{13} h^{-1} M_\odot < M < 10^{14} h^{-1} M_\odot$,
        The relative difference are shown in the bottom panel.
        For the FoF groups with
        $M > 10^{14} h^{-1} M_\odot$,
        there are about $1500$ gas partile
        in the nearest bin on the right of $10^6 K$.
        For the FoF groups with
        $10^{13} h^{-1} M_\odot < M < 10^{14} h^{-1} M_\odot$,
        there are about $4000$ gas particles
        in the nearest bin on the right of $10^4 K$.
        The unit of mass is $h^{-1}M_\odot$.
        }
    \label{fig:TPdf}
\end{figure}

\begin{figure}
    \includegraphics[width=\myfigwidtha]{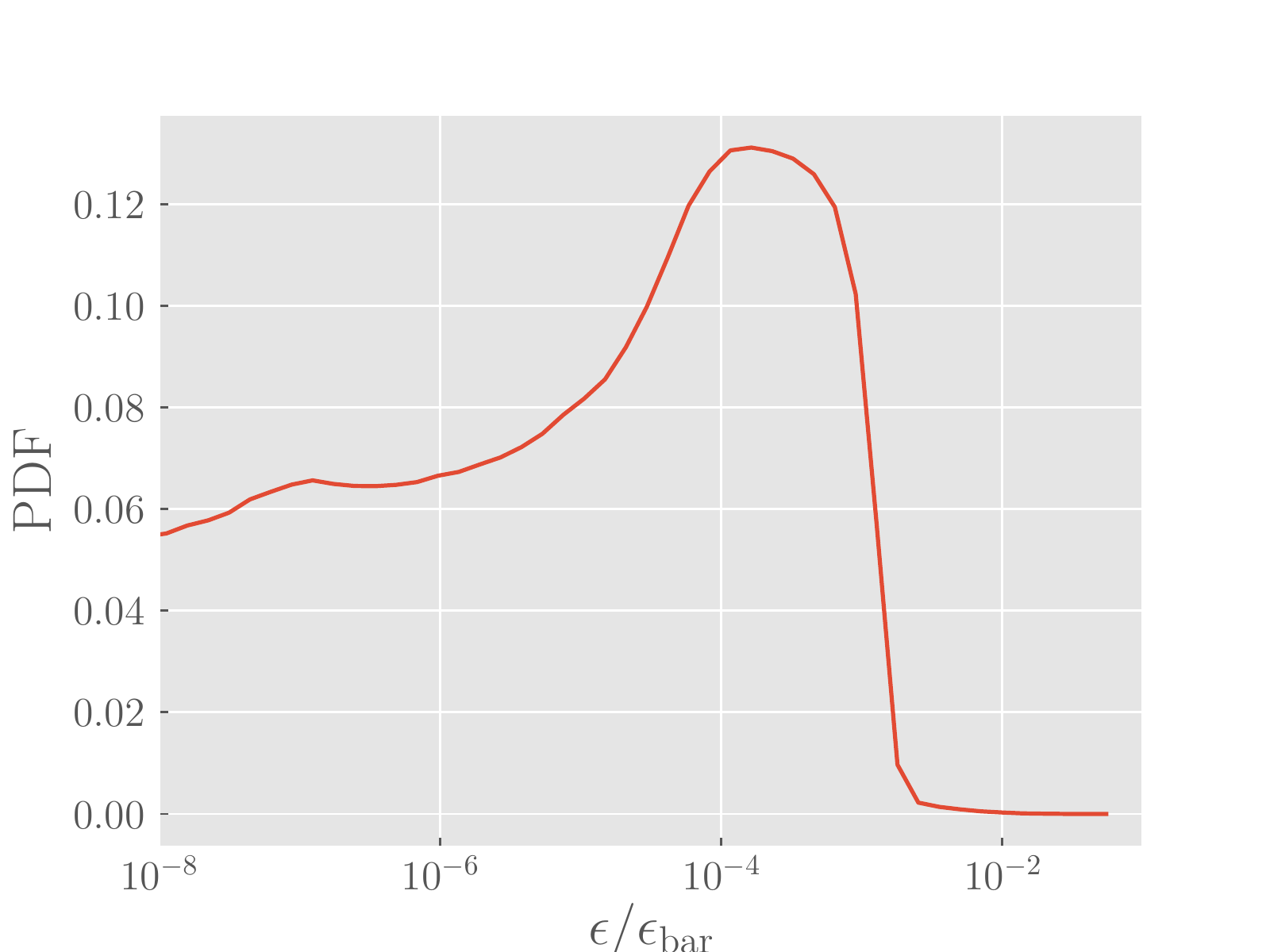}
    \caption{\
        PDF of the CRE energy of particles taken from
        the snapshot with reshift $z=0$
    }
    \label{fig:EPdf}
\end{figure}

\subsection{The impacts of CRE}
In this subsection, we investigate the impacts 
of the CRE processes on the cosmological hydro-simulation.

In Fig.~\ref{fig:ppdf}, we give 
Probability Density Function (PDF) of CRE pressure and CRP
pressure taken from the snapshot of SIM-CRE simulation with $z=0$.
For most of the gas particles,
the ratio of the CRP pressure to the total gas
pressure $P_{\rm CRP}/P_{\rm tot}$
($P_{\rm tot} = P_{\rm bar} + P_{\rm CRP} + P_{\rm CRE}$)
is about $0.1-1$, which indicates that the CRP pressure
is important to the hydro-simulation \citep{Jubelgas_2008}.
Since most gas particles with
$P_{\rm CRE} / P_{\rm tot} \lesssim 10^{-4}$,
the CRE pressure can be ignored in hydro-simulation.

In Fig.~\ref{fig:gaspha}, we present the $\rho-T$ phase-space
diagram (particle number in each bin) of
SIM simulation at $z=2$ and
$z=0$, the relative difference
(hereafter "difference") of phase-space diagram
between SIM and SIM-CRE are plotted at the bottom.
In order to avoid the statistical error,
we only plot the difference for the bins with
the number of particles larger than $10^4$.
As shown in the upper panel of Fig.~\ref{fig:gaspha},
several well-known features can be readily identified
\citep{Dave_2001, Vogelsberger_2012}:

    \textit{Diffuse}:
    ${\rho}/\bar{\rho} < 1000, T < 10^5 {\rm K}$.
    Photoionized intergalactic gas.

    \textit{Condensed}:
    ${\rho}/\bar{\rho} > 1000, T < 10^5{\rm K}$.
    Stars and cool galactic gas.

    \textit{Warm-hot}:
    $ 10^5 {\rm K} < T < 10^7 {\rm K}$.
    Warm-hot intergalactic medium.

    \textit{Hot}: $T > 10^7 {\rm K}$.
    Gas in galaxy clusters and large groups.

    Since the gas with $\rho/\bar{\rho} \sim 10-10^4$ and
    $T \sim 10^5-10^7 \rm {\rm K}$
    is comprised of the shock-heated gas 
    \mbox{\citep{Vogelsberger_2012}} and
    our injection source of CRE is only DSA,
    the difference of phase-space state may occur in \textit{warm-hot} phase,
    from right of Fig.\mbox{~\ref{fig:gaspha}},
    we lean that the influence  
    of the CRE processes on the \textit{warm-hot}
    intergalactic medium is up to $\pm 3\%$.
    The CRE processes does not affect the narrow ridge with
    $\rho/\bar{\rho}<10$ and $T<10^5 {\rm K}$
    in the diffuse photoionized intergalactic gas, the physics
    of which are only adiabatic expansion cooling and
    photoionization heating \citep{Vogelsberger_2012}.

In Fig.~\ref{fig:mf}, we plot the mass function of SIM and
SIM-CRE at z=0,
the difference of mass function between SIM and SIM-CRE
(middle panel),
and the number of group in each bin (bottom panel),
where we apply the FOF algorithm with a link length
parameter 0.16 to all particles
(Dark Matter, Gas and Star).
From the top panel of Fig.~\ref{fig:mf}, we find that the
influence of the CRE processes on mass function is up to $5\%$.
Since there are few number of group with 
$M > 10^{13}\, h^{-1} M_{\odot}$,
the difference of mass function above
$10^{13}\, h^{-1} M_{\odot}$ 
contain much statistical error. 

Finally, we investigate the impacts of the CRE processes on the 
gas temperature in FoF groups. We plot the 
PDF of the gas temperature in FoF groups and its 
the difference between our two simulations
in Fig.~\ref{fig:TPdf}, from which we find that
the CRE processes have a slight impact on the gas
temperature.
To give a relatively reliable result, we neglect
bins with an insufficient amount of gas particles (<$10^3$).
For the FoF groups with
$M>10^{14}\, h^{-1} M_\odot$,
the largest difference occurs 
on the right of $10^6 K$ with
a difference of $\sim 10\%$
caused by the CRE processes.
Similarly, for the FoF groups with
$10^{13} \, h^{-1} M_\odot < M < 10^{14} \, h^{-1} M_\odot$,
the influence of the CRE processes on the gas
temperature will reach about $5\%$.

To summarize, since the DSA injection efficiency
$\zeta_{\rm DSA}=0.005$,
the energy of CRE for most particles
are very small compared to the baryon energy 
(the ratio $\lesssim 0.1\%$, see the fourth column of
Fig.~\ref{fig:group} or Fig.~\ref{fig:EPdf}),
the accuracy of energy and pressure
of our method is $>96\%$ within $300 \, \rm Myr$  
(see Fig.~\ref{fig:comp}), which
guarantees that the result does not deviate too much.
However, the CRE processes can cause 
several percentage points of influence on hydro-simulation,
especially on the gas temperature of massive galaxy cluster
with $M > 10^{14} \, h^{-1} M_\odot$.

\section{Summary}\label{sec:sum}
In this paper, we have presented an approximative framework to
trace CRE physics and its implementation in hydro-simulation
which is capable of carrying out high-resolution simulations
of cosmological structure formation with CRE
physics.

In our method, we use a simplified power law for the momentum
distribution with spatially varying 
amplitude, upper cut-off, lower cut-off, and 
spectral index to approximate the real CRE spectrum
for each fluid element.  The on-the-fly shock detection
scheme for SPH developed by \citet{Pfrommer_2006} 
allows us to estimate Mach number, such that we can 
use DSA with an appropriate efficiency for CRE injection, 
and then we use the principles of conservation of
energy to derive the spectral parameters
after DSA injection. Coulomb cooling and radiative cooling 
mainly occur in low energy and high energy, respectively,
in order to account for these cooling and follow the evolution
of CRE spectrum after injection, we develop an approximating
method to integrate these losses, which
reach a balance between the complexity of CRE physics and
the requirement of computational efficiency and enable us to
determine the cut-offs of CRE spectrum. We also discuss the 
accuracy of our method by comparing with the numerical
solution of FP equation, the dynamical quantities like CRE
energy and pressure are reasonably well represented by
our method with an accuracy $>96\%$ within 
$\rm 300 Myr$ even if the number density
and the distribution at low energy does not match
which are unimportant for the use in hydro-simulation and
computation of radio emission. We also take
the adiabatic gains and losses into account.

\citet{Dolag_Stasyszyn_2009} have implemented MHD treatment in
SPH-simulation, which allows us to trace the magnetic field in an
MHD-simulation and then compute radio emission from the CRE spectrum.
The radio flux densities and spectral index for
the massive clusters in the simulation are in agreement with
observations \citep{Feretti_2012, Keshet_2004},
and the radio background of intergalactic
shocks estimated from our simulation is consistent with the
previous result \citep{Keshet_2004}. Our result also shows that
relaxed clusters have lower fluxes.

We have present the discussion about the impacts of
the CRE processes on the cosmological hydro-simulation.
We found that the CRE pressure can be ignored in hydro-simulation,
the phase-space diagram of gas is altered up to $3\%$
in \textit{warm-hot} phase,
and the influence of the CRE processes on the mass function
in the mass range $10^{12}-10^{13} h^{-1} M_\odot$ is up to $5\%$.
Finally, we discuss the impact of the CRE processes on
the gas temperature of the FoF group at $z=0$, and find
that the influence of the CRE processes on the gas temperature
of the FoF group with $M > 10^{14} h^{-1} M_\odot$ will reach
$\sim 10\%$.

\section*{Acknowledgements}
We are grateful to Volker Springel for his kind offer of
the developer version of the Gadget-3 code.
All simulations and analysis were performed on the
high-performance cluster at Center for Astronomy and
Astrophysics (CAA) at Shanghai Jiao Tong University.
This work is supported by
the Ministry of Science and Technology of China
(grant No. 2018YFA0404601),
the National Natural Science Foundation of China
(grant Nos. 11433002, 11621303, 61371147),
the National Key Research and Discovery Plan
(grant No. 2017YFF0210903),
and IBS under the project code, IBS-R018-D1.

\bibliographystyle{mnras}
\bibliography{references}

\begin{thebibliography}{}
\makeatletter
\relax
\def\mn@urlcharsother{\let\do\@makeother \do\$\do\&\do\#\do\^\do\_\do\%\do\~}
\def\mn@doi{\begingroup\mn@urlcharsother \@ifnextchar [ {\mn@doi@}
  {\mn@doi@[]}}
\def\mn@doi@[#1]#2{\def\@tempa{#1}\ifx\@tempa\@empty \href
  {http://dx.doi.org/#2} {doi:#2}\else \href {http://dx.doi.org/#2} {#1}\fi
  \endgroup}
\def\mn@eprint#1#2{\mn@eprint@#1:#2::\@nil}
\def\mn@eprint@arXiv#1{\href {http://arxiv.org/abs/#1} {{\tt arXiv:#1}}}
\def\mn@eprint@dblp#1{\href {http://dblp.uni-trier.de/rec/bibtex/#1.xml}
  {dblp:#1}}
\def\mn@eprint@#1:#2:#3:#4\@nil{\def\@tempa {#1}\def\@tempb {#2}\def\@tempc
  {#3}\ifx \@tempc \@empty \let \@tempc \@tempb \let \@tempb \@tempa \fi \ifx
  \@tempb \@empty \def\@tempb {arXiv}\fi \@ifundefined
  {mn@eprint@\@tempb}{\@tempb:\@tempc}{\expandafter \expandafter \csname
  mn@eprint@\@tempb\endcsname \expandafter{\@tempc}}}

\bibitem[\protect\citeauthoryear{{Balsara}}{{Balsara}}{1995}]{Balsara_1995}
{Balsara} D.~S.,  1995, \mn@doi [Journal of Computational Physics]
  {10.1016/S0021-9991(95)90221-X}, \href
  {https://ui.adsabs.harvard.edu/abs/1995JCoPh.121..357B} {121, 357}

\bibitem[\protect\citeauthoryear{{Brunetti} \& {Lazarian}}{{Brunetti} \&
  {Lazarian}}{2011}]{Brunetti_2011}
{Brunetti} G.,  {Lazarian} A.,  2011, \mn@doi [\mnras]
  {10.1111/j.1365-2966.2010.17457.x}, \href
  {http://adsabs.harvard.edu/abs/2011MNRAS.410..127B} {410, 127}

\bibitem[\protect\citeauthoryear{{Brunetti}, {Blasi}, {Cassano}  \&
  {Gabici}}{{Brunetti} et~al.}{2004}]{Brunetti_2004}
{Brunetti} G.,  {Blasi} P.,  {Cassano} R.,   {Gabici} S.,  2004, \mn@doi
  [\mnras] {10.1111/j.1365-2966.2004.07727.x}, \href
  {http://adsabs.harvard.edu/abs/2004MNRAS.350.1174B} {350, 1174}

\bibitem[\protect\citeauthoryear{{Bryan} et~al.,}{{Bryan} et~al.}{2014}]{Enzo}
{Bryan} G.~L.,  et~al., 2014, \mn@doi [\apjs] {10.1088/0067-0049/211/2/19},
  \href {http://adsabs.harvard.edu/abs/2014ApJS..211...19B} {211, 19}

\bibitem[\protect\citeauthoryear{{Buote}}{{Buote}}{2001}]{Buote_2001}
{Buote} D.~A.,  2001, \mn@doi [\apjl] {10.1086/320500}, \href
  {https://ui.adsabs.harvard.edu/abs/2001ApJ...553L..15B} {553, L15}

\bibitem[\protect\citeauthoryear{{Cassano} \& {Brunetti}}{{Cassano} \&
  {Brunetti}}{2005}]{Cassano_2005}
{Cassano} R.,  {Brunetti} G.,  2005, \mn@doi [\mnras]
  {10.1111/j.1365-2966.2005.08747.x}, \href
  {http://adsabs.harvard.edu/abs/2005MNRAS.357.1313C} {357, 1313}

\bibitem[\protect\citeauthoryear{{Chang} \& {Cooper}}{{Chang} \&
  {Cooper}}{1970}]{Chang_1970}
{Chang} J.~S.,  {Cooper} G.,  1970, \mn@doi [Journal of Computational Physics]
  {10.1016/0021-9991(70)90001-X}, \href
  {http://adsabs.harvard.edu/abs/1970JCoPh...6....1C} {6, 1}

\bibitem[\protect\citeauthoryear{{Crocce}, {Pueblas}  \&
  {Scoccimarro}}{{Crocce} et~al.}{2006}]{2lpt_2006}
{Crocce} M.,  {Pueblas} S.,   {Scoccimarro} R.,  2006, \mn@doi [\mnras]
  {10.1111/j.1365-2966.2006.11040.x}, \href
  {https://ui.adsabs.harvard.edu/abs/2006MNRAS.373..369C} {373, 369}

\bibitem[\protect\citeauthoryear{{Dav{\'e}} et~al.,}{{Dav{\'e}}
  et~al.}{2001}]{Dave_2001}
{Dav{\'e}} R.,  et~al., 2001, \mn@doi [\apj] {10.1086/320548}, \href
  {http://adsabs.harvard.edu/abs/2001ApJ...552..473D} {552, 473}

\bibitem[\protect\citeauthoryear{{Dedner}, {Kemm}, {Kr{\"o}ner}, {Munz},
  {Schnitzer}  \& {Wesenberg}}{{Dedner} et~al.}{2002}]{Dedner_2002}
{Dedner} A.,  {Kemm} F.,  {Kr{\"o}ner} D.,  {Munz} C.-D.,  {Schnitzer} T.,
  {Wesenberg} M.,  2002, \mn@doi [Journal of Computational Physics]
  {10.1006/jcph.2001.6961}, \href
  {https://ui.adsabs.harvard.edu/abs/2002JCoPh.175..645D} {175, 645}

\bibitem[\protect\citeauthoryear{{Dolag} \& {Stasyszyn}}{{Dolag} \&
  {Stasyszyn}}{2009}]{Dolag_Stasyszyn_2009}
{Dolag} K.,  {Stasyszyn} F.,  2009, \mn@doi [\mnras]
  {10.1111/j.1365-2966.2009.15181.x}, \href
  {http://adsabs.harvard.edu/abs/2009MNRAS.398.1678D} {398, 1678}

\bibitem[\protect\citeauthoryear{{Donnert} \& {Brunetti}}{{Donnert} \&
  {Brunetti}}{2014}]{Donnert_Brunetti_2014}
{Donnert} J.,  {Brunetti} G.,  2014, \mn@doi [\mnras] {10.1093/mnras/stu1417},
  \href {http://adsabs.harvard.edu/abs/2014MNRAS.443.3564D} {443, 3564}

\bibitem[\protect\citeauthoryear{{Efstathiou}, {Bond}  \& {White}}{{Efstathiou}
  et~al.}{1992}]{Efstathiou_1992}
{Efstathiou} G.,  {Bond} J.~R.,   {White} S.~D.~M.,  1992, \mn@doi [\mnras]
  {10.1093/mnras/258.1.1P}, \href
  {http://adsabs.harvard.edu/abs/1992MNRAS.258P...1E} {258, 1P}

\bibitem[\protect\citeauthoryear{{En{\ss}lin}, {Pfrommer}, {Springel}  \&
  {Jubelgas}}{{En{\ss}lin} et~al.}{2007}]{Ensslin_2007}
{En{\ss}lin} T.~A.,  {Pfrommer} C.,  {Springel} V.,   {Jubelgas} M.,  2007,
  \mn@doi [\aap] {10.1051/0004-6361:20065294}, \href
  {http://adsabs.harvard.edu/abs/2007A%26A...473...41E} {473, 41}

\bibitem[\protect\citeauthoryear{{Feretti}, {Giovannini}, {Govoni}  \&
  {Murgia}}{{Feretti} et~al.}{2012}]{Feretti_2012}
{Feretti} L.,  {Giovannini} G.,  {Govoni} F.,   {Murgia} M.,  2012, \mn@doi
  [\aapr] {10.1007/s00159-012-0054-z}, \href
  {http://adsabs.harvard.edu/abs/2012A%26ARv..20...54F} {20, 54}

\bibitem[\protect\citeauthoryear{{Fermi}}{{Fermi}}{1949}]{Fermi_1949}
{Fermi} E.,  1949, \mn@doi [Physical Review] {10.1103/PhysRev.75.1169}, \href
  {http://adsabs.harvard.edu/abs/1949PhRv...75.1169F} {75, 1169}

\bibitem[\protect\citeauthoryear{{Hoeft} \& {Br{\"u}ggen}}{{Hoeft} \&
  {Br{\"u}ggen}}{2007}]{Hoeft_Bruggen_2007}
{Hoeft} M.,  {Br{\"u}ggen} M.,  2007, \mn@doi [\mnras]
  {10.1111/j.1365-2966.2006.11111.x}, \href
  {http://adsabs.harvard.edu/abs/2007MNRAS.375...77H} {375, 77}

\bibitem[\protect\citeauthoryear{{Hoeft}, {Br{\"u}ggen}, {Yepes},
  {Gottl{\"o}ber}  \& {Schwope}}{{Hoeft} et~al.}{2008}]{Hoeft_2008}
{Hoeft} M.,  {Br{\"u}ggen} M.,  {Yepes} G.,  {Gottl{\"o}ber} S.,   {Schwope}
  A.,  2008, \mn@doi [\mnras] {10.1111/j.1365-2966.2008.13955.x}, \href
  {http://adsabs.harvard.edu/abs/2008MNRAS.391.1511H} {391, 1511}

\bibitem[\protect\citeauthoryear{{Jubelgas}, {Springel}  \& {Dolag}}{{Jubelgas}
  et~al.}{2004}]{Jubelgas_2004}
{Jubelgas} M.,  {Springel} V.,   {Dolag} K.,  2004, \mn@doi [\mnras]
  {10.1111/j.1365-2966.2004.07801.x}, \href
  {http://adsabs.harvard.edu/abs/2004MNRAS.351..423J} {351, 423}

\bibitem[\protect\citeauthoryear{{Jubelgas}, {Springel}, {En{\ss}lin}  \&
  {Pfrommer}}{{Jubelgas} et~al.}{2008}]{Jubelgas_2008}
{Jubelgas} M.,  {Springel} V.,  {En{\ss}lin} T.,   {Pfrommer} C.,  2008,
  \mn@doi [\aap] {10.1051/0004-6361:20065295}, \href
  {http://adsabs.harvard.edu/abs/2008A%26A...481...33J} {481, 33}

\bibitem[\protect\citeauthoryear{{Katz}, {Weinberg}  \& {Hernquist}}{{Katz}
  et~al.}{1996}]{Katz_1996}
{Katz} N.,  {Weinberg} D.~H.,   {Hernquist} L.,  1996, \mn@doi [\apjs]
  {10.1086/192305}, \href {http://adsabs.harvard.edu/abs/1996ApJS..105...19K}
  {105, 19}

\bibitem[\protect\citeauthoryear{{Keshet}, {Waxman}  \& {Loeb}}{{Keshet}
  et~al.}{2004}]{Keshet_2004}
{Keshet} U.,  {Waxman} E.,   {Loeb} A.,  2004, \mn@doi [\apj] {10.1086/424837},
  \href {http://adsabs.harvard.edu/abs/2004ApJ...617..281K} {617, 281}

\bibitem[\protect\citeauthoryear{{Landau} \& {Lifshitz}}{{Landau} \&
  {Lifshitz}}{1959}]{Landau_1959}
{Landau} L.~D.,  {Lifshitz} E.~M.,  1959, {Fluid mechanics}

\bibitem[\protect\citeauthoryear{{Lawson}, {Mayer}, {Osborne}  \&
  {Parkinson}}{{Lawson} et~al.}{1987}]{Lawson_1987}
{Lawson} K.~D.,  {Mayer} C.~J.,  {Osborne} J.~L.,   {Parkinson} M.~L.,  1987,
  \mn@doi [\mnras] {10.1093/mnras/225.2.307}, \href
  {https://ui.adsabs.harvard.edu/abs/1987MNRAS.225..307L} {225, 307}

\bibitem[\protect\citeauthoryear{{Longair}}{{Longair}}{2011}]{Longair_2011}
{Longair} M.~S.,  2011, {High Energy Astrophysics}

\bibitem[\protect\citeauthoryear{{Marinacci}, {Vogelsberger}, {Mocz}  \&
  {Pakmor}}{{Marinacci} et~al.}{2015}]{Marinacci_2015}
{Marinacci} F.,  {Vogelsberger} M.,  {Mocz} P.,   {Pakmor} R.,  2015, \mn@doi
  [\mnras] {10.1093/mnras/stv1692}, \href
  {http://adsabs.harvard.edu/abs/2015MNRAS.453.3999M} {453, 3999}

\bibitem[\protect\citeauthoryear{{Marinacci} et~al.,}{{Marinacci}
  et~al.}{2018}]{Marinacci_2018}
{Marinacci} F.,  et~al., 2018, \mn@doi [\mnras] {10.1093/mnras/sty2206}, \href
  {http://adsabs.harvard.edu/abs/2018MNRAS.480.5113M} {480, 5113}

\bibitem[\protect\citeauthoryear{{Mignone} \& {Tzeferacos}}{{Mignone} \&
  {Tzeferacos}}{2010}]{Mignone_2010}
{Mignone} A.,  {Tzeferacos} P.,  2010, \mn@doi [Journal of Computational
  Physics] {10.1016/j.jcp.2009.11.026}, \href
  {https://ui.adsabs.harvard.edu/abs/2010JCoPh.229.2117M} {229, 2117}

\bibitem[\protect\citeauthoryear{{Monaghan}}{{Monaghan}}{1992}]{Monaghan_1992}
{Monaghan} J.~J.,  1992, \mn@doi [\araa] {10.1146/annurev.aa.30.090192.002551},
  \href {http://adsabs.harvard.edu/abs/1992ARA%26A..30..543M} {30, 543}

\bibitem[\protect\citeauthoryear{{Monaghan}}{{Monaghan}}{2005}]{Monaghan_2005}
{Monaghan} J.~J.,  2005, \mn@doi [Reports on Progress in Physics]
  {10.1088/0034-4885/68/8/R01}, \href
  {http://adsabs.harvard.edu/abs/2005RPPh...68.1703M} {68, 1703}

\bibitem[\protect\citeauthoryear{{Pakmor}, {Bauer}  \& {Springel}}{{Pakmor}
  et~al.}{2011}]{Pakmor_2011}
{Pakmor} R.,  {Bauer} A.,   {Springel} V.,  2011, \mn@doi [\mnras]
  {10.1111/j.1365-2966.2011.19591.x}, \href
  {https://ui.adsabs.harvard.edu/abs/2011MNRAS.418.1392P} {418, 1392}

\bibitem[\protect\citeauthoryear{{Park} \& {Petrosian}}{{Park} \&
  {Petrosian}}{1995}]{Park_1995}
{Park} B.~T.,  {Petrosian} V.,  1995, \mn@doi [\apj] {10.1086/175828}, \href
  {http://adsabs.harvard.edu/abs/1995ApJ...446..699P} {446, 699}

\bibitem[\protect\citeauthoryear{{Park} \& {Petrosian}}{{Park} \&
  {Petrosian}}{1996}]{Park_1996}
{Park} B.~T.,  {Petrosian} V.,  1996, \mn@doi [\apjs] {10.1086/192278}, \href
  {http://adsabs.harvard.edu/abs/1996ApJS..103..255P} {103, 255}

\bibitem[\protect\citeauthoryear{{Petkova} \& {Springel}}{{Petkova} \&
  {Springel}}{2009}]{Petkova_2009}
{Petkova} M.,  {Springel} V.,  2009, \mn@doi [\mnras]
  {10.1111/j.1365-2966.2009.14843.x}, \href
  {http://adsabs.harvard.edu/abs/2009MNRAS.396.1383P} {396, 1383}

\bibitem[\protect\citeauthoryear{{Pfrommer}, {Springel}, {En{\ss}lin}  \&
  {Jubelgas}}{{Pfrommer} et~al.}{2006}]{Pfrommer_2006}
{Pfrommer} C.,  {Springel} V.,  {En{\ss}lin} T.~A.,   {Jubelgas} M.,  2006,
  \mn@doi [\mnras] {10.1111/j.1365-2966.2005.09953.x}, \href
  {http://adsabs.harvard.edu/abs/2006MNRAS.367..113P} {367, 113}

\bibitem[\protect\citeauthoryear{{Pinzke}, {Oh}  \& {Pfrommer}}{{Pinzke}
  et~al.}{2017}]{Pinzke_2017}
{Pinzke} A.,  {Oh} S.~P.,   {Pfrommer} C.,  2017, \mn@doi [\mnras]
  {10.1093/mnras/stw3024}, \href
  {http://adsabs.harvard.edu/abs/2017MNRAS.465.4800P} {465, 4800}

\bibitem[\protect\citeauthoryear{{Rybicki} \& {Lightman}}{{Rybicki} \&
  {Lightman}}{1979}]{Rybicki_Lightman_1979}
{Rybicki} G.~B.,  {Lightman} A.~P.,  1979, {Radiative processes in
  astrophysics}

\bibitem[\protect\citeauthoryear{{Scholz} \& {Walters}}{{Scholz} \&
  {Walters}}{1991}]{Scholz_1991}
{Scholz} T.~T.,  {Walters} H.~R.~J.,  1991, \mn@doi [\apj] {10.1086/170587},
  \href {http://adsabs.harvard.edu/abs/1991ApJ...380..302S} {380, 302}

\bibitem[\protect\citeauthoryear{{Sijacki}, {Springel}, {Di Matteo}  \&
  {Hernquist}}{{Sijacki} et~al.}{2007}]{Blackhole2}
{Sijacki} D.,  {Springel} V.,  {Di Matteo} T.,   {Hernquist} L.,  2007, \mn@doi
  [\mnras] {10.1111/j.1365-2966.2007.12153.x}, \href
  {http://adsabs.harvard.edu/abs/2007MNRAS.380..877S} {380, 877}

\bibitem[\protect\citeauthoryear{{Springel}}{{Springel}}{2005}]{Gadget2}
{Springel} V.,  2005, \mn@doi [\mnras] {10.1111/j.1365-2966.2005.09655.x},
  \href {http://adsabs.harvard.edu/abs/2005MNRAS.364.1105S} {364, 1105}

\bibitem[\protect\citeauthoryear{{Springel}}{{Springel}}{2010}]{moving-mesh_2010}
{Springel} V.,  2010, \mn@doi [\mnras] {10.1111/j.1365-2966.2009.15715.x},
  \href {http://adsabs.harvard.edu/abs/2010MNRAS.401..791S} {401, 791}

\bibitem[\protect\citeauthoryear{{Springel}}{{Springel}}{2011a}]{springel_2011}
{Springel} V.,  2011a, arXiv e-prints, \href
  {http://adsabs.harvard.edu/abs/2011arXiv1109.2218S} {}

\bibitem[\protect\citeauthoryear{{Springel}}{{Springel}}{2011b}]{moving-mesh_2011}
{Springel} V.,  2011b, arXiv e-prints, \href
  {http://adsabs.harvard.edu/abs/2011arXiv1109.2218S} {}

\bibitem[\protect\citeauthoryear{{Springel} \& {Hernquist}}{{Springel} \&
  {Hernquist}}{2003}]{Springel_Hernquist_2003}
{Springel} V.,  {Hernquist} L.,  2003, \mn@doi [\mnras]
  {10.1046/j.1365-8711.2003.06206.x}, \href
  {http://adsabs.harvard.edu/abs/2003MNRAS.339..289S} {339, 289}

\bibitem[\protect\citeauthoryear{{Springel}, {Yoshida}  \& {White}}{{Springel}
  et~al.}{2001}]{Gadget1}
{Springel} V.,  {Yoshida} N.,   {White} S.~D.~M.,  2001, \mn@doi [\na]
  {10.1016/S1384-1076(01)00042-2}, \href
  {http://adsabs.harvard.edu/abs/2001NewA....6...79S} {6, 79}

\bibitem[\protect\citeauthoryear{{Springel}, {Di Matteo}  \&
  {Hernquist}}{{Springel} et~al.}{2005a}]{Blackhole1}
{Springel} V.,  {Di Matteo} T.,   {Hernquist} L.,  2005a, \mn@doi [\mnras]
  {10.1111/j.1365-2966.2005.09238.x}, \href
  {http://adsabs.harvard.edu/abs/2005MNRAS.361..776S} {361, 776}

\bibitem[\protect\citeauthoryear{{Springel} et~al.,}{{Springel}
  et~al.}{2005b}]{Millennium}
{Springel} V.,  et~al., 2005b, \mn@doi [\nat] {10.1038/nature03597}, \href
  {http://adsabs.harvard.edu/abs/2005Natur.435..629S} {435, 629}

\bibitem[\protect\citeauthoryear{{Stasyszyn}, {Dolag}  \& {Beck}}{{Stasyszyn}
  et~al.}{2013}]{Stasyszyn_2013}
{Stasyszyn} F.~A.,  {Dolag} K.,   {Beck} A.~M.,  2013, \mn@doi [\mnras]
  {10.1093/mnras/sts018}, \href
  {https://ui.adsabs.harvard.edu/abs/2013MNRAS.428...13S} {428, 13}

\bibitem[\protect\citeauthoryear{{Vazza}, {Dolag}, {Ryu}, {Brunetti},
  {Gheller}, {Kang}  \& {Pfrommer}}{{Vazza} et~al.}{2011}]{Vazza_2011}
{Vazza} F.,  {Dolag} K.,  {Ryu} D.,  {Brunetti} G.,  {Gheller} C.,  {Kang} H.,
   {Pfrommer} C.,  2011, \mn@doi [\mnras] {10.1111/j.1365-2966.2011.19546.x},
  \href {https://ui.adsabs.harvard.edu/abs/2011MNRAS.418..960V} {418, 960}

\bibitem[\protect\citeauthoryear{{Vogelsberger}, {Sijacki}, {Kere{\v s}},
  {Springel}  \& {Hernquist}}{{Vogelsberger} et~al.}{2012}]{Vogelsberger_2012}
{Vogelsberger} M.,  {Sijacki} D.,  {Kere{\v s}} D.,  {Springel} V.,
  {Hernquist} L.,  2012, \mn@doi [\mnras] {10.1111/j.1365-2966.2012.21590.x},
  \href {http://adsabs.harvard.edu/abs/2012MNRAS.425.3024V} {425, 3024}

\bibitem[\protect\citeauthoryear{{Vogelsberger} et~al.,}{{Vogelsberger}
  et~al.}{2014}]{Illustris}
{Vogelsberger} M.,  et~al., 2014, \mn@doi [\nat] {10.1038/nature13316}, \href
  {http://adsabs.harvard.edu/abs/2014Natur.509..177V} {509, 177}

\bibitem[\protect\citeauthoryear{{Waxman} \& {Loeb}}{{Waxman} \&
  {Loeb}}{2000}]{waxman_2000}
{Waxman} E.,  {Loeb} A.,  2000, \mn@doi [\apjl] {10.1086/317326}, \href
  {http://adsabs.harvard.edu/abs/2000ApJ...545L..11W} {545, L11}

\makeatother
\end{thebibliography}

\appendix
\section{Formula}\label{sec:appd}
The incomplete Beta function is
\begin{equation}
    B_{x}\left(a,b\right) = \int_{0}^{x} t^{a-1} \left(1-t\right)^{b-1} dt.
\end{equation}
We introduce the symbol $B_{mn}$, i.e.
\begin{equation}
    B_{mn} = \frac{1}{2}B_{\frac{1}{1+q^2}}\left(\frac{\alpha-m}{2},\frac{n-\alpha}{2}\right).
\end{equation}
The relations between the dimensionless velocity, the Lorentz factor and the dimensionless momemtum are
\begin{equation}
    \beta = \frac{p}{\sqrt{1+p^2}}, \quad \gamma_L = \sqrt{1+p^2}.
\end{equation}
So
\begin{equation}
    \frac{d \gamma}{dp} = \beta, \quad \frac{d \beta}{dp} = \frac{1}{\gamma^3}.
\end{equation}
Defining $t = \frac{1}{1+p^2}$, we get
\begin{equation}
    \begin{split}
        &p = \left(\frac{1-t}{t}\right)^{\frac{1}{2}} \\
        &\frac{dp}{dt} = -\frac{1}{2} \frac{t^{-\frac{3}{2}}}{\sqrt{1-t}},
    \end{split}
\end{equation}
from which we can do the following useful integrals,
\begin{equation}
    \begin{split}
        &\int_{q}^{\infty} \frac{p^{-\alpha}}{\sqrt{1+p^2}} = \int_{\frac{1}{1+q^2}}^{0}
        \left(\frac{1-t}{t}\right)^{\frac{-\alpha}{2}} t^{\frac{1}{2}}
        \left(-\frac{1}{2} \frac{t^{-\frac{3}{2}}}{\sqrt{1-t}}\right) dt \\
        &= \frac{1}{2} \int_{0}^{\frac{1}{1+q^2}} t^{\frac{\alpha}{2} + \frac{1}{2} -
        \frac{3}{2}} \left(1-t\right)^{\frac{-\alpha}{2} - \frac{1}{2}} dt \\
        &= \frac{1}{2} \int_{0}^{\frac{1}{1+q^2}} t^{\frac{\alpha}{2}-1}
        \left(1-t\right)^{\frac{1-\alpha}{2} - 1} dt \\
        &= \frac{1}{2}B_{01},
    \end{split}
\end{equation}
and

\begin{equation}
    \begin{split}
        &\int_{q}^{\infty} \sqrt{1+p^2} p^{-\alpha} dp = \int_{q}^{\infty}
        \sqrt{1+p^2} \frac{dp^{{1-\alpha}}}{1-\alpha}  \\
        &= \frac{1}{\alpha-1} q^{1-\alpha} \sqrt{1+q^2} \\
        &- \int_{q}^{\infty} \frac{p^{2-\alpha}}{(1-\alpha)\sqrt{1+p^2}} \\
        &= \frac{1}{\alpha-1}\left[ \frac{1}{2} B_{23} + q^{1-\alpha}
        \sqrt{1+q^2} \right].
    \end{split}
\end{equation}
So the CRE energy $\epsilon$ and pressure $P$ are given by
\begin{equation}
    \begin{split}
        &\epsilon = \int_{q}^{\infty} m c^2 \left(\sqrt{1+p^2}
        - 1\right) \quad C p^{-\alpha} dp \\
        &= \frac{C m c^2}{\alpha-1}\left[\frac{1}{2}B_{23}
        + q^{1-\alpha}\left(\sqrt{1+q^2} - 1\right)\right],
    \end{split}
\end{equation}
    \begin{equation}
        \begin{split}
            &P = \int_{q}^{\infty} \frac{m_ec^2 \beta p}{3} \quad C p^{-\alpha} dp \\
            &= \frac{C m c^2}{3}\int_{q}^{\infty} \frac{p^{2-\alpha}}{\sqrt{1+p^2}} dp \\
            &= \frac{C m c^2}{6} B_{23}.
        \end{split}
    \end{equation}

\section{Visualization}\label{sec:visual}
\begin{figure*}
    \includegraphics[width=\myfigwidthb]{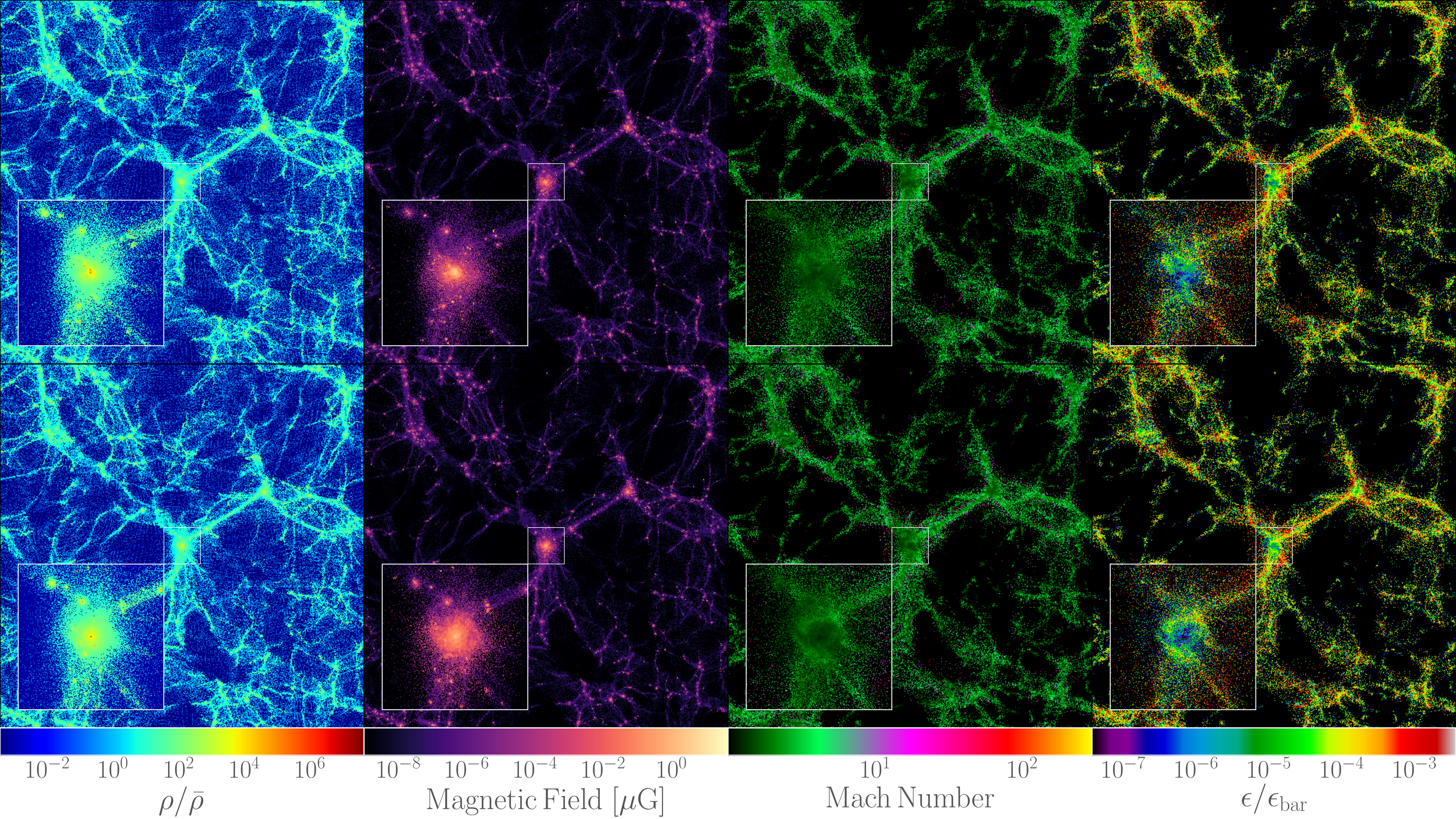}
    \caption{Visulization of SIM-CRE at redshift $z=0.1$ (top panels)
    and $z=0$ (bottom panels). These pictures have a comoving side length of
    $\rm 100\,h^{-1}Mpc$ while the projection length along the line of sight
    amounts to $10 \, h^{-1} \rm Mpc$. The zoom-in plot extents $10 \, h^{-1} \rm Mpc$  and
    contains the most massive cluster in slimulation.
    }
    \label{fig:slices}
\end{figure*}

\bsp
\label{lastpage}

\end{document}